\documentclass[showpacs,preprintnumbers,amsmath,amssymb,aps,prb,twocolumn,superscriptaddress,10pt]{revtex4-1}
\usepackage{amsfonts}
\usepackage{graphicx}
\usepackage{color}
\usepackage{bm}

\usepackage{caption,subcaption}
\usepackage{epstopdf}

\usepackage{subfig}

\renewcommand{\i}{\mathrm{i}}
\newcommand{\e}{\mathrm{e}}

\renewcommand{\Re}{\mathrm{Re}}
\renewcommand{\Im}{\mathrm{Im}}

\begin{document}

\title{Switching and propagation of magneto-plasmon-polaritons in magnetic slot waveguides and cavities}
\author{D.~Nikolova}
\affiliation{London Centre for Nanotechnology, University College London, 17-19 Gordon Street, London WC1H 0AH, U.K.}
\email{dessie.nikolova@gmail.com}
\author{ A.J.~Fisher}
\affiliation{London Centre for Nanotechnology, University College London, 17-19 Gordon Street, London WC1H 0AH, U.K.}
\affiliation{UCL Department of Physics and Astronomy, University College London, Gower Street, London WC1E 6BT, U.K.}

\begin{abstract}
The dispersion relations for surface plasmon-polaritons propagating in the Voigt geometry in a metal-insulator-metal waveguide with a magneto-optically active dielectric medium are derived.  The symmetry between the upper and lower interfaces is broken by the introduction of the magnetic field; the balance between the field distributions on the two interfaces can be controlled by the applied field.  This control is illustrated by finite-element method numerical simulations of the field distributions around a point dipole placed in the centre of the short waveguide; it is shown that both the total emission of radiation from the cavity and the distribution of the far-field radiation can be strongly modified by tuning the magnetisation of the waveguide. This raises the novel possibility of using magnetic fields to control light propagation in nanostructures.
\end{abstract}

\pacs{42.79.Gn,85.70.Sq,73.20.Mf}

\maketitle

\section{Introduction}
Surface plasmon-polaritons (SPPs) \cite{RaetherSPP} are electromagnetic waves that propagate along metal-dielectric interfaces and can be guided by metallic nanostructures beyond the diffraction limit \cite{PlasmonicsReview,MaierPlasmonicsReviews}.  They are attractive because of the very small length scales over which it is possible to localise the electromagnetic fields, giving very wide scope for the manipulation of those fields on the nanoscale. They usually extend few hundreds of nanometers in the surrounding dielectric. They can be further confined and guided in metal-insulator-metal (MIM) structures, which are building blocks for resonant guided wave networks  \cite{AtwaterNetworks} and can be used for future on-chip nanocircuits \cite{BrongersmaNanocurcuits,BrongersmaScience}.

Volume modes in waveguides propagate via multiple reflections from the metal surfaces and the energy is localised in the core of the structure. Surface plasmon modes, on the other hand, are localised at the interfaces; for a structure with two symmetric interfaces they couple to produce symmetric or antisymmetric mode profiles. 
The case of metal-insulator-metal (MIM)  plasmonic waveguides with a non-magnetic dielectric has been studied previously both theoretically \cite{MiMWavegudeIMI,MIMWaveguide,ModesMIMTheory,ModalAnalysisMIM} and experimentally \cite{MIM,DispMIMExperimental,MIMExperimental}. In these studies, MIM plasmons were excited by an external light source using, for example, in-coupling through slits in one of the metal cladding layers, or via the electron beam of a scanning electron microscope.
 
Placing an emitter inside a MIM cavity is another possible route to excite cavity plasmons.   Dipolar emitters placed in MIM slab and slot waveguide structures were found to couple strongly to the plasmon waveguide modes \cite{DipoleInMIM}.    Such a structure can serve as an optical nanoantenna \cite{PlasmonicBeaming}.
 Active control of the direction of emission of an optical nanoantenna has been achieved by electrically controlling its load impedance \cite{ElectricalTunningAntenna}. The electric field has been shown \cite{ElopticalModulation}  to modulate  the dispersion relation of the plasmons propagating at the interface of an electro-optically active dielectric and metal.

A quite different route for active control of plasmons is by using a magnetic field. The combination of plasmonics with magneto-optical materials is particularly interesting because it introduces a nanoscale interaction between light fields and magnetisation, hence opening up the possibility of using either one of these fields to control the other.  For example, external magnetic fields could then be used to control plasmonic devices through nanoscale analogies of the Kerr and Faraday effects\cite{FreiserMagnetooptic}; alternatively, the light field could be used to interact with a nanoscale magnetic system.  SPPs propagating at a single metal-dielectric  interface where one or both media are magnetic have been investigated theoretically and experimentally \cite{HaleviVoigt,ChiuQuinn,NonreciprocalSPPtheory,MPNanostructures, magnetoplasmonsExp}. Plasmons at the interface between a magnetic metal and a photonics crystal have also been considered \cite{NonreciprocialFan}. Enhanced magneto-optical activity occurs as a result of  surface plasmons in nanodisks with a magnetic metal \cite{MOactivityMPnanodisks,EnhancedMOactivity,EnhancedMOactivityAPL,NickelNanoantennas,DesignerMagnetoplasmons} or magnetic dielectric \cite{SuPREMO}.

One notable effect occurs in the Voigt geometry (magnetisation perpendicular to the propagation direction), where the wave vectors for left and right propagation become unequal.   There has been growing interest in these ideas in recent years, and also in experimentally realising active magnetic control of the plasmons.   For example, in \cite{ActiveMagnetoPlasmonics} it was demonstrated that a d.c. magnetic field can be used to modulate the plasmons, and it has also been shown that introducing a periodic nanostructure results in an enhanced magneto-optical Kerr effect (MOKE) signal \cite{MOKEprl,MagnetoPlasmonicCrystal}. For a review on magneto-plasmonics see \cite{MagnetoPlasmonicsReview, AOMmagnetoplasmonicsReview}.  
Magneto-plasmonic waveguides consisting of  insulator-metal-insulator (IMI) geometry with  a ferromagnetic metal surrounded by nonmagnetic dielectrics \cite{LongRangeSPPVoigt} or a nonmagnetic metal bounded by
ferromagnetic dielectrics \cite{MOeffectSPPwaveguides} have been studied for application as active devices in SPP-based optics.   MIM cavities with magnetic metals have been studied in \cite{MIMmagneticMetal} where  it was observed that the magnetic modulation of SPP is higher when only one of the metallic interfaces is magnetic. 

In this article we  consider the case of a MIM waveguide containing a magnetic insulator. We study the surface plasmon modes of the waveguide, a preliminary investigation of which was published in \cite{DessieSPIE}. We then employ numerical simulation to show the field and energy distribution inside the waveguide. We excite the modes by placing a dipole in the middle of the structure and show that  an external magnetic field can switch on and off the coupling of the dipole to the surface plasmons. We study the out-coupled radiation, the intensity and direction of which can also be controlled through external magnetic field. 
\section{Magneto-plasmon dispersion relations}

\subsection{Single magnetic interface}\label{sec:singleint}
To understand the behaviour of the SPP guiding modes in a cavity we first review the case of a single interface separating a metal and a magnetic dielectric, which we take to be isotropic apart from the magnetic effects.  We take the boundary at the plane $y=0$, with the wave propagating in the $x$-direction and the static magnetisation being in the $z$ direction (i.e. in the Voigt geometry). For the Voigt geometry only TM modes can propagate on this interface \cite{Landau}, so we also take the oscillating $\mathbf{H}$-field in the $z$-direction. The permittivity tensor for a gyrotropic medium when $\mathbf{M}=(0,0,M_z)$  is
\begin{equation}
\left(
 \begin{array}{ccc}

        \epsilon_r & \epsilon_{xy}  & 0 \\
      -\epsilon_{xy}& \epsilon_r & 0  \\
      0 & 0 & \epsilon_r 
 \end{array} \right)
 \label{dieltensor}
 \end{equation}
 with $\epsilon_{xy}=gM_z$.
Let the half-space $y>0$ contain the magnetic dielectric and $y<0$ the non-magnetic metal. We seek to find waves damped as $y \rightarrow\pm\infty$ in the form
\begin{equation}
\begin{split}
H_d=&H_0\e^{\i k_xx-\kappa_{yd}y}, \\
\kappa_{yd}=&\sqrt{k_x^2-w^2\epsilon_d/c^2},  \quad y\ >\ 0 \nonumber
\end{split}
\end{equation}
in the dielectric, and
\begin{equation}
\begin{split}
H_m=&H_0\e^{\i k_xx+\kappa_{ym}y}, \\
 \kappa_{ym}=&\sqrt{k_x^2-w^2\epsilon_m/c^2}, \quad y\ <\ 0
\end{split}
\label{kymeqn}
\end{equation}
in the metal, where $\epsilon_d=\epsilon_{r}+\epsilon_{xy}^2/\epsilon_{r}$ and $\epsilon_m$ is the permittivity of the metal. 

The dispersion relation can be found from the boundary conditions on the tangential $\mathbf{E}$ and normal $\mathbf{D}$ fields and is given by  
\begin{equation}\label{eq:conditionsingle}
\epsilon_{m}\epsilon_r \kappa_{yd}-\mathrm{i} \epsilon_{m}\epsilon_{xy} k_{x} + (\epsilon_{r}^2+\epsilon_{xy}^2) \kappa_{ym}=0,
\end{equation}
which for the nonmagnetic case $\epsilon_{xy}=0$ results in the well-known plasmon dispersion relation $k_x=\pm k_0\sqrt{\frac{\epsilon_m  \epsilon_r}{(  \epsilon_m +  \epsilon_r)}}$, where $k_0=\omega/c$ is the magnitude of the free-space wave vector corresponding to angular frequency $\omega$, in the region where $\epsilon_m  \epsilon_r<0$ and $\epsilon_m +  \epsilon_r<0$. 


To first order in $\epsilon_{xy}$, the solution for $k_x$ at a magnetic interface is
\begin{equation}
k_x=+k_+\quad\mbox{or}\quad -k_-
\end{equation}
with
\begin{equation}
k_\pm=\sqrt{\frac{\epsilon_m  \epsilon_r}{(  \epsilon_m +  \epsilon_r)}}\left(1 \pm \frac{\sqrt{( \epsilon_r -  \epsilon_m)^2  \epsilon_r\epsilon_m^5} \epsilon_{xy}}{ ( \epsilon_r -  \epsilon_m)^2  \epsilon_r\epsilon_m ( \epsilon_m +  \epsilon_r)}\right) k_0
\label{kxSingle}
\end{equation}
Correspondingly, for the decay constants in the $y$ direction one obtains the solutions
\begin{eqnarray} 
\kappa_{yd}=-\frac{\epsilon_{r}\kappa_{ym}}{\epsilon_m} +\frac{\i \epsilon_{xy}k_x}{\epsilon_r}\\
\kappa_{ym}=-\frac{\epsilon_{m}\kappa_{yd}}{\epsilon_r} +\frac{\i \epsilon_m \epsilon_{xy}k_x}{\epsilon_r^2}
\label{SingleIFky}
\end{eqnarray}
Note in particular the structure of equation~(\ref{kxSingle}): the right-moving ($\Re(k_x)>0$) and left-moving ($\Re(k_x)<0$) solutions have different magnitudes, because the left-right symmetry is broken by the introduction of the magnetic field.  
The symmetry is only restored if the sign of the static magnetic field is also reversed.  For $\Im(\epsilon_{xy})>0$ we have $\Re(k_+)>\Re(k_-)$, and \textit{vice versa}.  This `non-reciprocal' behaviour has been observed experimentally in surface plasmon propagation with a magnetic medium \cite{MSPPpropagationExp, NonreciprocialSpoof}. Note also how the presence of a real (dissipative) part in $\epsilon_{xy}$, introducing a non-Hermitian part to (\ref{dieltensor}) contributes to the imaginary part of $k_x$ and hence increases the attenuation of the wave, while a purely imaginary  $\epsilon_{xy}$ leaves (\ref{dieltensor}) Hermitian and introduces no additional dissipation.  

\subsection{Metal-Insulator-Metal waveguide}
\begin{figure}[t] 
   \centering
 \includegraphics[width=0.88\columnwidth]{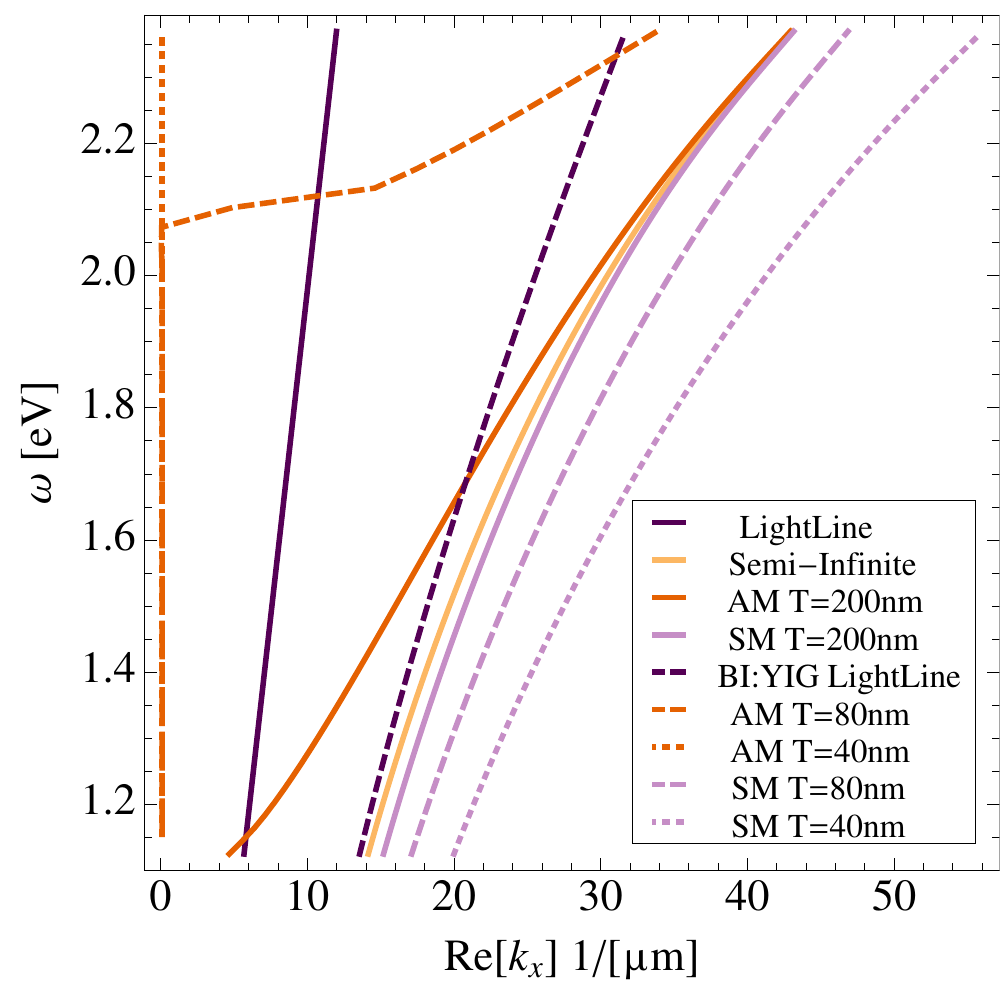} 

 \caption{The solutions of the dispersion relation Eq. (\ref{MIMdisp}) for $\omega$ as a function of $k_x$ for the non-magnetic case for a range of cavity sizes. The thick blue line is the solution for a single interface. The coupled cavity modes are split into low-energy symmetric (SM) and high-energy anti-symmetric (AM) modes. The black lines are the light lines in free space (thick) and in the dielectric (dashed). }
   \label{DispRelation}
\end{figure}

Consider now the geometry shown in figure \ref{DispRelation}(a).  We seek to find plasmon solutions, i.e. surface-bound  waves that in the metal are of the form 
\begin{equation}
H_m=H_0\e^{\i k_xx-\i\omega t+\kappa_{ym}T/2}\e^{-\kappa_{ym}|y|}, \quad |y| >T/2
\label{First}
\end{equation}
and in the dielectric are linear combinations of decaying exponentials from either side
\begin{equation}
H_{d}=(A\e^{\kappa_{yd}y}+B\e^{-\kappa_{yd}y})\e^{\i k_xx-\i\omega t},\quad  |y| <T/2.
\label{Hdiel}
\end{equation}
In general $|A|\neq |B|$ when the dielectric is anisotropic (i.e. when $\epsilon_{xy}\ne 0$). 

From the Maxwell equations the electric fields in the metal have the form
\begin{eqnarray} 
  E_{xm}&=& \frac{-\i \kappa_{ym} }{\omega\epsilon_m}\frac{|y|}{y}H_m \nonumber \\
  E_{ym}&=& \frac{k_{x} }{\omega\epsilon_m}H_m
  \label{Emetal}
\end{eqnarray}
while the $E$-fields in the dielectric for an $H$-field of the form (\ref{Hdiel}) are
\begin{widetext}
\begin{eqnarray} 
  E_{xd}= -\frac{\e^{\i k_xx-\i\omega t} \left[\i\epsilon_r\kappa_{yd}(B\e^{-\kappa_{yd}y}-A\e^{\kappa_{yd}y})+\epsilon_{xy}k_x (A\e^{\kappa_{yd}y}+B\e^{-\kappa_{yd}y})\right]}{\omega (\epsilon_{r}^2+\epsilon_{xy}^2)} \nonumber \\
 E_{yd}=\frac{\e^{\i k_xx-i\omega t}\left[\i\epsilon_rk_{x}(-B\e^{-\kappa_{yd}y}-A\e^{\kappa_{yd}y})+\epsilon_{xy}\kappa_{yd} (A\e^{\kappa_{yd}y}-B\e^{-\kappa_{yd}y})\right]}{\omega (\epsilon_{r}^2+\epsilon_{xy}^2)}.
    \label{antiE}
\end{eqnarray}
From the boundary conditions at the two interfaces at $y=\pm T/2$ we obtain two equations for $A$ and $B$. In order for them to have non-zero solutions the determinant formed from the coefficients should equal zero. This condition gives the dispersion relation for the system:
\begin{equation}
\frac{(2\epsilon_r \epsilon_m(\epsilon_{r}^2+\epsilon_{xy}^2) \kappa_{yd}\kappa_{ym} \cosh[\kappa_{yd}T]+(\epsilon_m^2(\epsilon_{xy}^2k_x^2+\epsilon_{r}^2 \kappa_{yd}^2) +(\epsilon_{r}^2+\epsilon_{xy}^2)^2\kappa_{ym}^2)\sinh[\kappa_{yd}T]}{\omega^2 \epsilon_m^2 (\epsilon_{r}^2+\epsilon_{xy}^2)^2}=0.
\label{MIMdisp}
\end{equation}
The solutions are 
\begin {equation}
k_x=\sqrt{\frac{\epsilon_m w^2}{c^2} +\frac{ \left(\epsilon_r \epsilon_m k_{yd} \coth[ k_{yd} T] \pm
     \epsilon_m \sqrt{(\epsilon_r^2 k_{yd}^2 -
         \epsilon_{xy}^2 (k_{yd}^2 + (\epsilon_r + \epsilon_{xy}^2/\epsilon_r) k_0^2) \sinh^2 [k_{yd} T])}\right)^2}{(\epsilon_r^2 + \epsilon_{xy}^2)^2 \sinh^2[k_{yd}T]}}
           \label{kxSols}
\end{equation}

In the limit of small $\epsilon_{xy}$ and small cavity thickness $T$ the solution can be approximated by 
\begin {equation}
k_x=\sqrt{\frac{\epsilon_m ^2 k_{yd}^2 \tanh[k_{yd} T/2]}{\epsilon_r^2}+ \epsilon_m \frac{w^2}{c^2}}+ \frac{\epsilon_{xy}^2 \epsilon_m^2 (- k_{yd}^2 + 
   \epsilon_r k_0^2 + (k_{yd}^2 + \epsilon_r k_0^2) \cosh[ k_{yd} T]) \tanh^2[
  k_{yd} T/2]}{2  \epsilon_r^4\sqrt{\frac{\epsilon_m ^2 k_{yd}^2 \tanh[k_{yd} T/2]}{\epsilon_r^2}+ \epsilon_m \frac{w^2}{c^2}}}
  \label{kxSolsSym}
  \end{equation}

\begin {equation}
k_x=\sqrt{\frac{\epsilon_m ^2 k_{yd}^2\coth[k_{yd} T/2]}{\epsilon_r^2}+\epsilon_m \frac{w^2}{c^2}}
+\frac{\epsilon_{xy}^2 \epsilon_m^2(-k_{yd}^2 + 
   \epsilon_r k_0^2 - (k_{yd}^2 + \epsilon_r k_0^2) \cosh[ k_{yd} T]) \coth^2[k_{yd} T/2]}{2 \epsilon_r^4\sqrt{\frac{\epsilon_m ^2 k_{yd}^2\coth[k_{yd} T/2]}{\epsilon_r^2}+\epsilon_m \frac{w^2}{c^2}}}
     \label{kxSolsAnti}
\end{equation}
\end{widetext}
Note the quadratic dependence on $\epsilon_{xy}$ in contrast with the linear dependence exhibited by equation (\ref{kxSingle}).

The corresponding values of $\kappa_{ym}$ can be immediately recovered from equation (\ref{kymeqn}).  
In the non-magnetic case  $\epsilon_{xy}=0$ we recover  the well-known (cf. \cite{MIM}) plasmon dispersion relations $\kappa_{ym1}=\frac{\epsilon_m\kappa_{yd}\tanh[\kappa_{yd}T/2]}{\epsilon_r}$
corresponding to the symmetric ($A=B$) and $\kappa_{ym2}=\frac{\epsilon_m\kappa_{yd}\coth[\kappa_{yd}T/2]}{\epsilon_r}$
to the antisymmetric ($A=-B$) $H$-fields respectively.  Note that in the non-magnetic limit the $E_x$ field has the opposite symmetry to $H_z$: $E_x$ is odd if $A=B$ and even if $A=-B$.  In this paper we shall always refer to the symmetry (or approximate symmetry) of the $H$-field when describing a mode.

  \begin{figure}[ht!] 
 {  \centering
 \includegraphics[width=0.6\columnwidth]{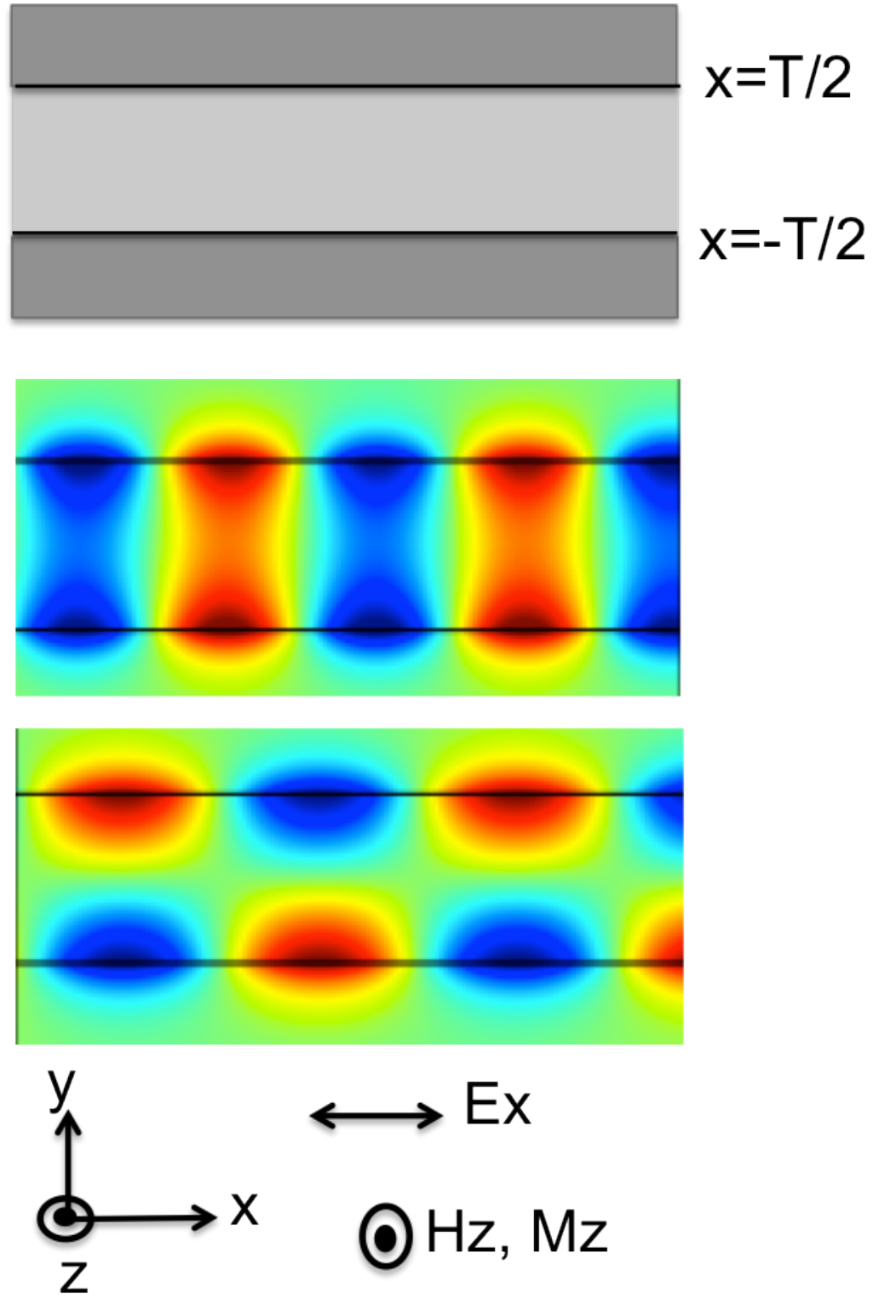} }
 \vspace{1mm}
 \caption{The MIM waveguide geometry. The coordinate system, showing the direction of the static magnetisation and plots of the electric field distribution for the antisymmetric (above) and symmetric (below) modes---note the symmetry descriptions refer to the symmetry of the $H$-field (not shown);}
   \label{Geometry}
\end{figure}


\begin{figure}[t!]
\begin{tabular}{cc}
 \begin{subfigure}[b]{0.45\columnwidth}
(a)  \includegraphics[width= \columnwidth]{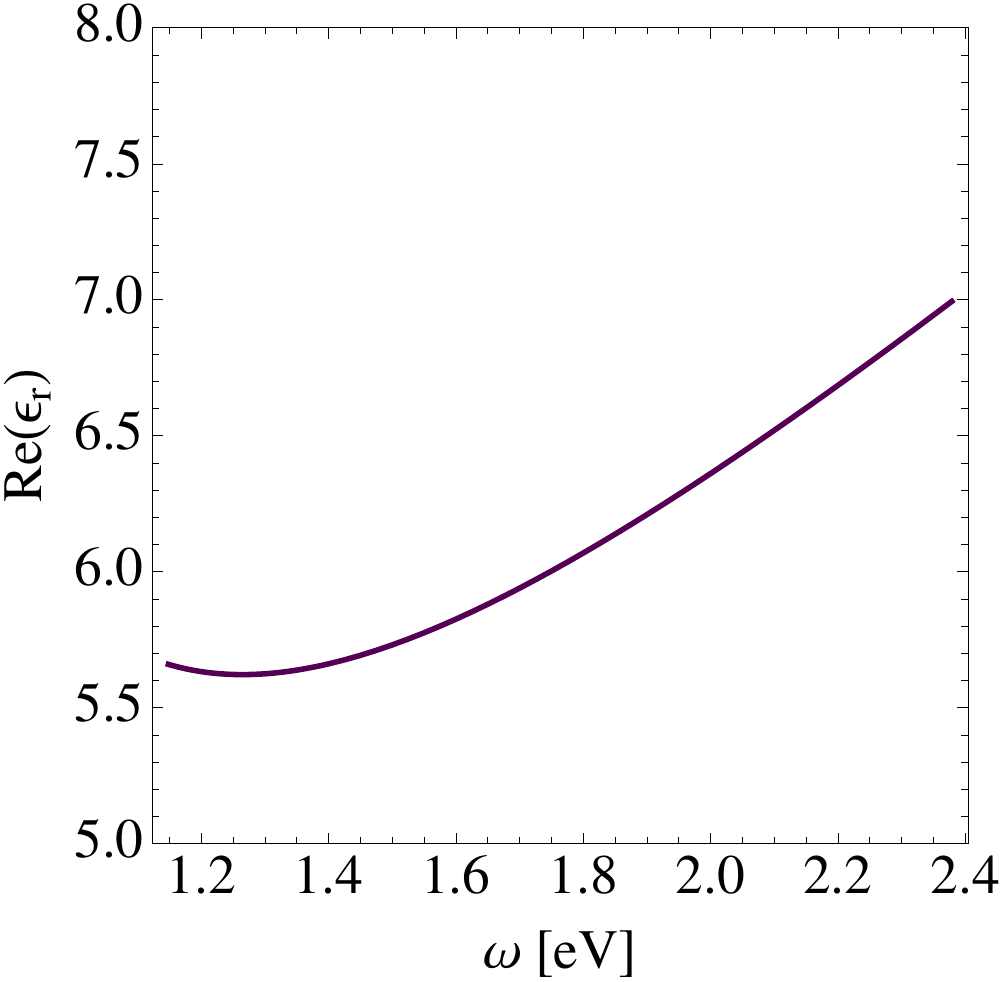} 
\end{subfigure}&
 \begin{subfigure}[b]{0.45 \columnwidth}
(b)  \includegraphics[width= \columnwidth]{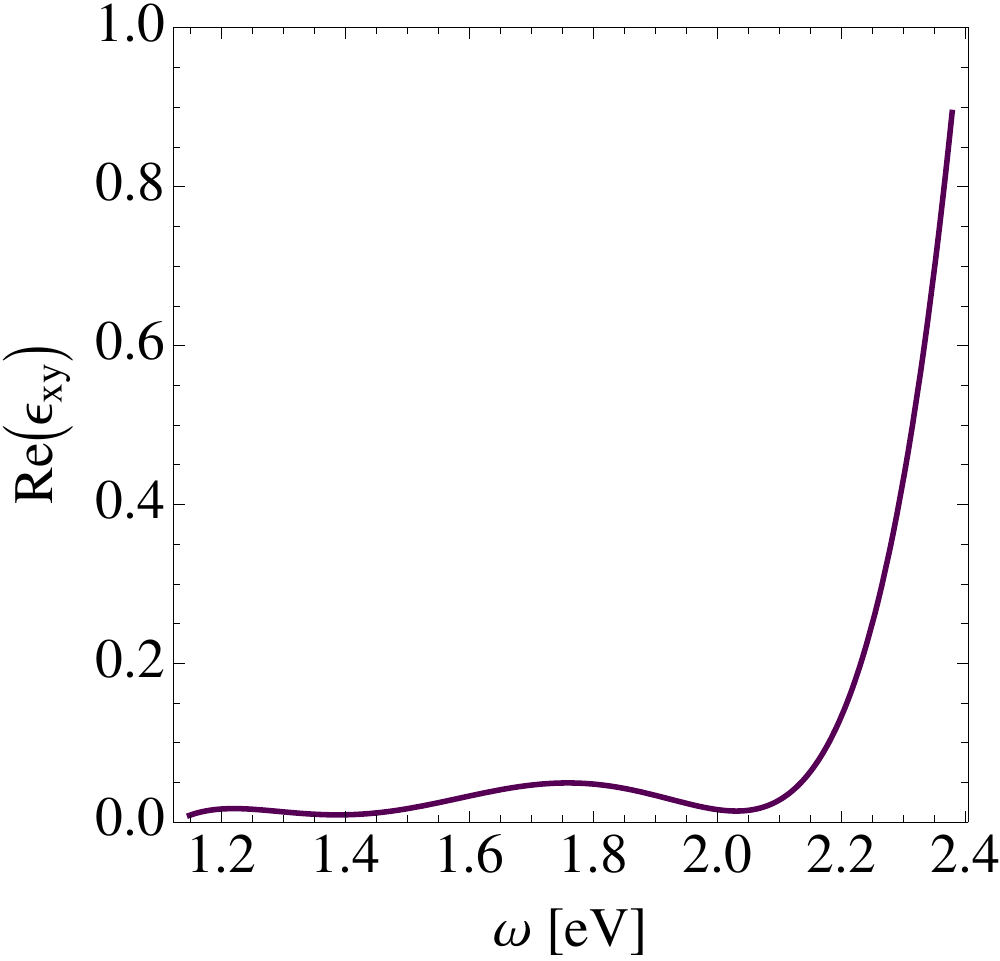} 
\end{subfigure}\\
 \begin{subfigure}[b]{0.45 \columnwidth}
(c)  \includegraphics[width= \columnwidth]{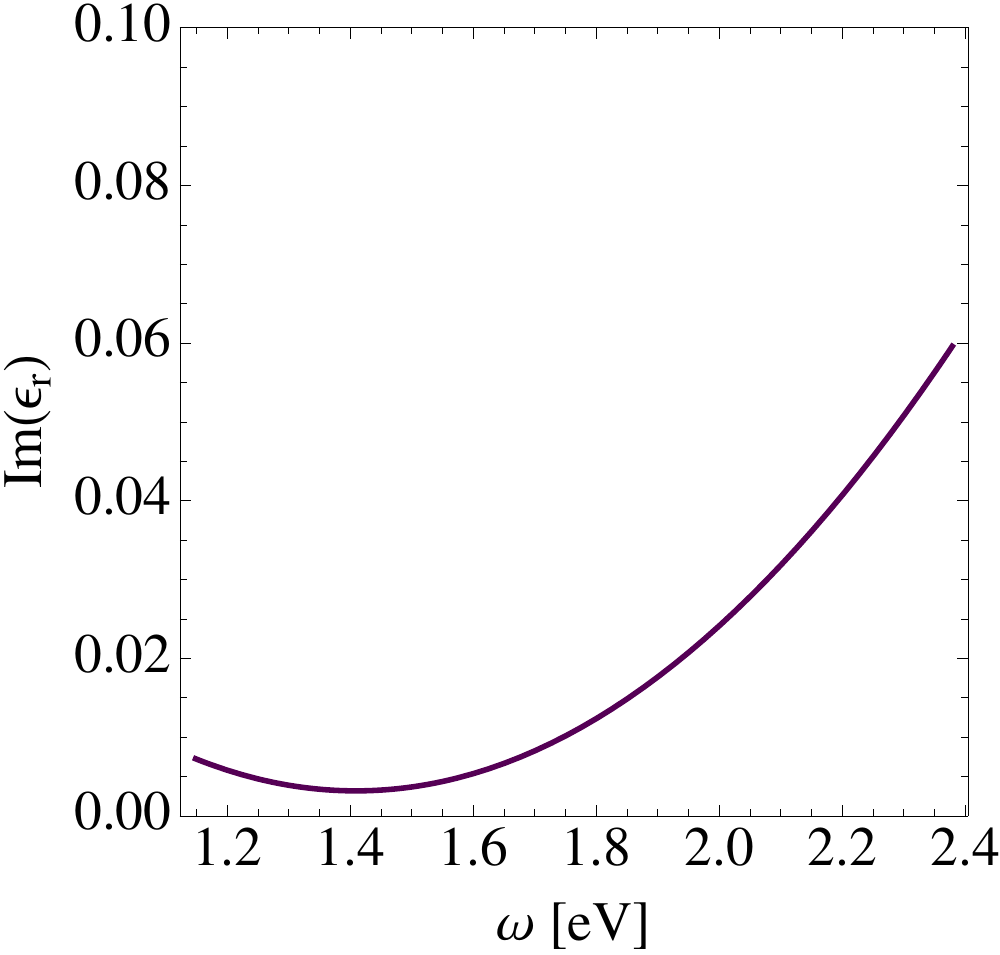} 
\end{subfigure}&
 \begin{subfigure}[b]{0.45 \columnwidth}
(d)  \includegraphics[width= \columnwidth]{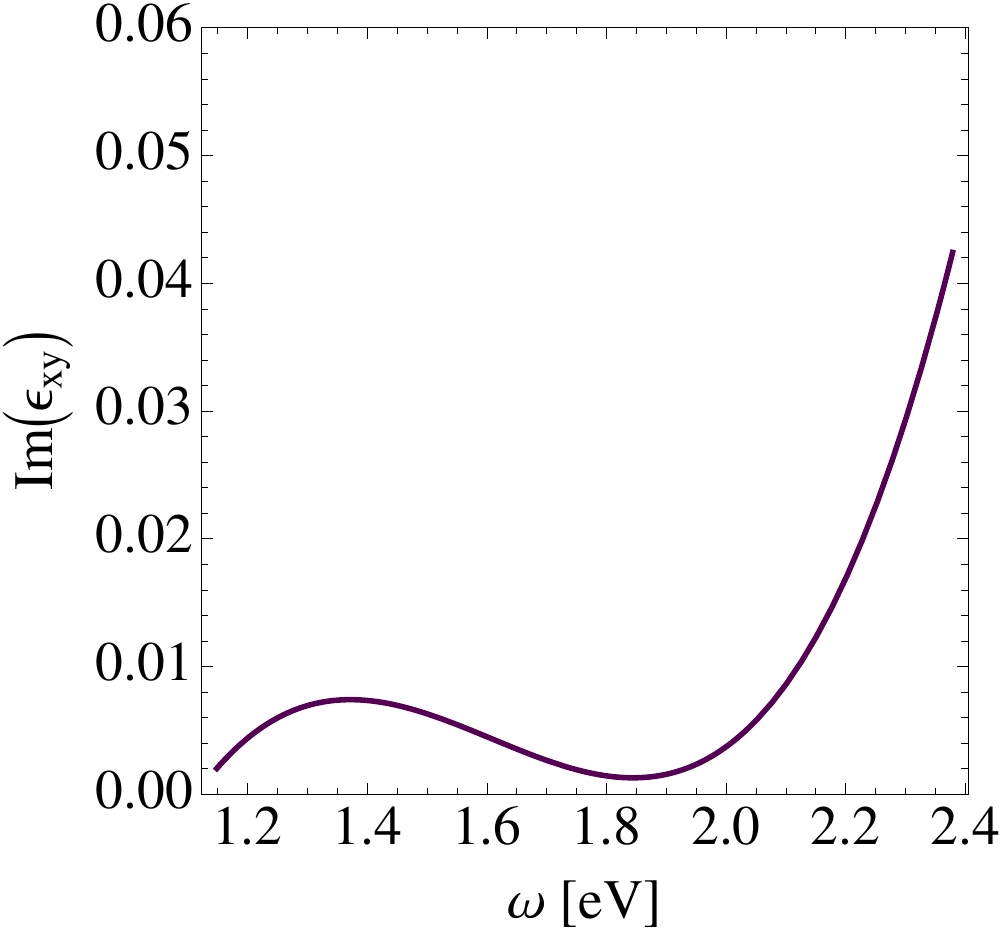} 
\end{subfigure}
\end{tabular}
\caption{Frequency dependence of the dielectric functions: real parts of (a) $\epsilon_r$  (b) $\epsilon_{xy}$; imaginary parts of (c)  $\epsilon_r$  and (d) $\epsilon_{xy}$.\label{dielectricfns}}

\end{figure}
For large cavities $\cosh[k_{yd} T] $ and $\sinh[k_{yd} T] $ in  the dispertion relation Eq. (\ref{MIMdisp}) approach the same value and we recover the single-interface solutions given by Eqs. (\ref{kxSingle})-(\ref{SingleIFky}).  There are two forward-propagating and two backward-propagating modes: the right-moving mode with $k_x=k_+$ and the left-moving mode with $k_x=-k_-$ are localised on the upper interface (which has the same orientation as that considered in \S\ref{sec:singleint}), while the other two modes  with $k_x=k_-$ and $k_x=-k_+$ are localised on the lower interface, which has the opposite orientation.  The additional degeneracy between positive and negative $k$ appears because the structure is now invariant under a reflection in the line $y=0$.

\subsection{Example system}
To illustrate these effects, we describe a specific example where the magnetic dielectric is bismuth-substituted yttrium-iron garnet (Bi:YIG) and the metal is silver. The magneto optical data for the Bi:YIG is taken from experimental data \cite{BiYIGdata,BiYIGdata2}. The permittivity for the silver is fitted from the Drude model 
\begin{equation}\epsilon_m=1-\frac{\omega_p^2}{\omega^2}+\i\frac{\omega_p^2\tau}{\omega(1+\omega^2\tau^2)}
\end{equation}
where $\omega_p$ is the bulk plasma frequency, 
with parameters taken from references \citenum{JohnsonChristy} and \citenum{SilverPlasmaFreq}. Both permittivities are therefore frequency dependent, as shown in Figure~\ref{dielectricfns}. Throughout we take $\hbar=1$ and so give values for angular frequency in energy units.
The geometry is shown in Figure \ref{Geometry}, which also shows the $E_x$ field distribution for a symmetric and anti-symmetric mode. Figure \ref{DispRelation} shows the dispersion relation of the waveguide for different thicknesses when there is no magnetisation in the dielectric. The thick blue line is the dispersion relation (Eq. \ref{eq:conditionsingle}) of surface plasmons on a single interface; for the MIM waveguide the mode frequency is split into two modes moving in each direction, as expected.  The lower lying cavity plasmonic modes have a symmetric magnetic field profile, while the high energy ones shown with the dark red line on the figure  are antisymmetric. For thicker cavities the energy difference between the symmetric and antisymmetric solutions gets smaller as one would expect as in the limit of infinite waveguide thickness both solutions will become almost degenerate, corresponding to the separate solutions (Eq. \ref{kxSingle}) on the upper and lower interfaces of the slab.  
 The thick black line is the vacuum light line $\omega=ck_x$;  solutions to the left of it are inaccessible to coupling from the outside world in a translationally invariant system. The dashed black line indicates the light line in the dielectric medium $\omega=ck_x/\sqrt{\epsilon_d}$; modes to the left of the dielectric light line have propagating solutions in the dielectric (i.e. nearly imaginary $\kappa_{yd}$, provided dissipation in the dielectric is weak).   Modes to the right of the dielectric light line have nearly real $\kappa_{yd}$; these are the bound surface-plasmon solutions that we seek.  The lowest symmetric mode always has a frequency below the plasmon of the semi-infinite surface and remains bound for arbitrarily small cavities; the lowest antisymmetric solution always spills out and becomes a propagating state in the dielectric region (i.e. crosses the light line of the dielectric) for sufficiently small $k_x$.  All modes are rapidly decaying into the metallic boundaries of the cavity.
\begin{figure*}[th] 
\begin{tabular}{cc}
 \begin{subfigure}[b]{0.47 \textwidth}
 (a) \\
   \centering
 \includegraphics[width= 0.9\textwidth]{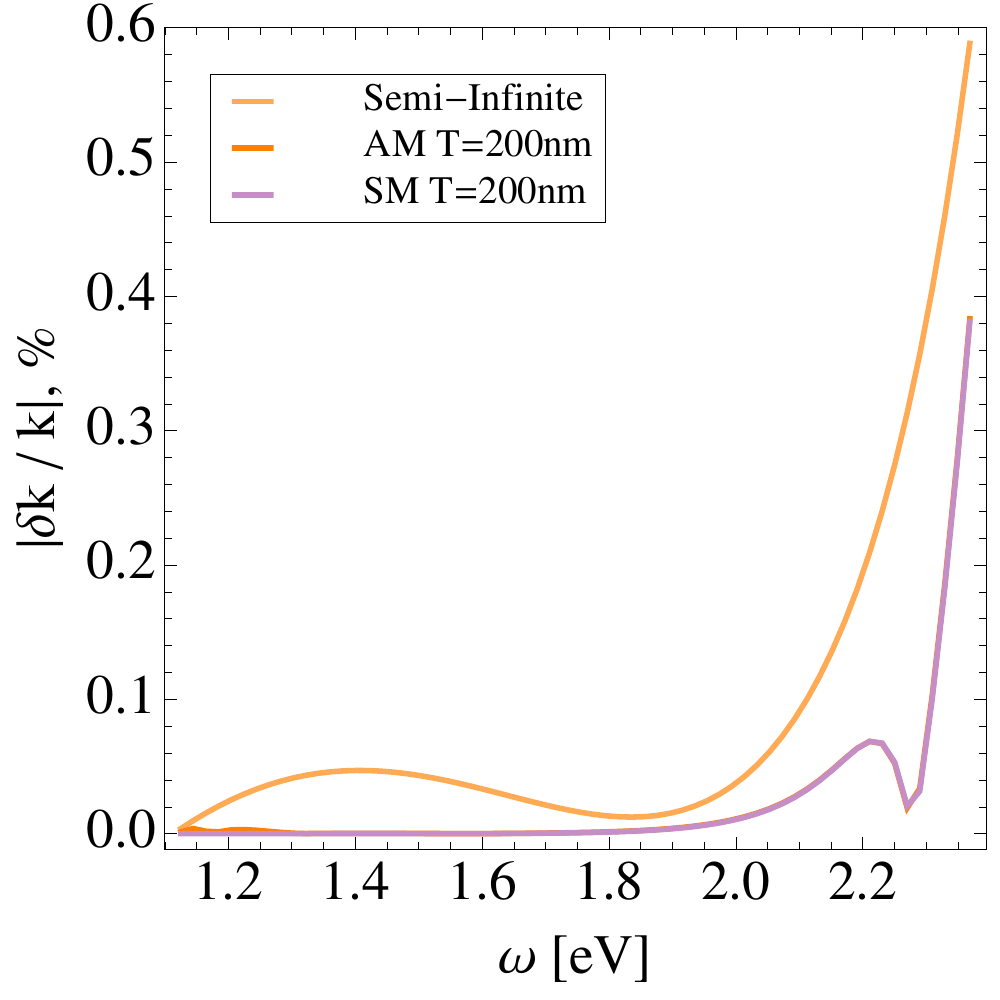} 
\end{subfigure}
 \begin{subfigure}[b]{0.47 \textwidth}
 (b)\\
   \centering
 \includegraphics[width= 0.88\textwidth]{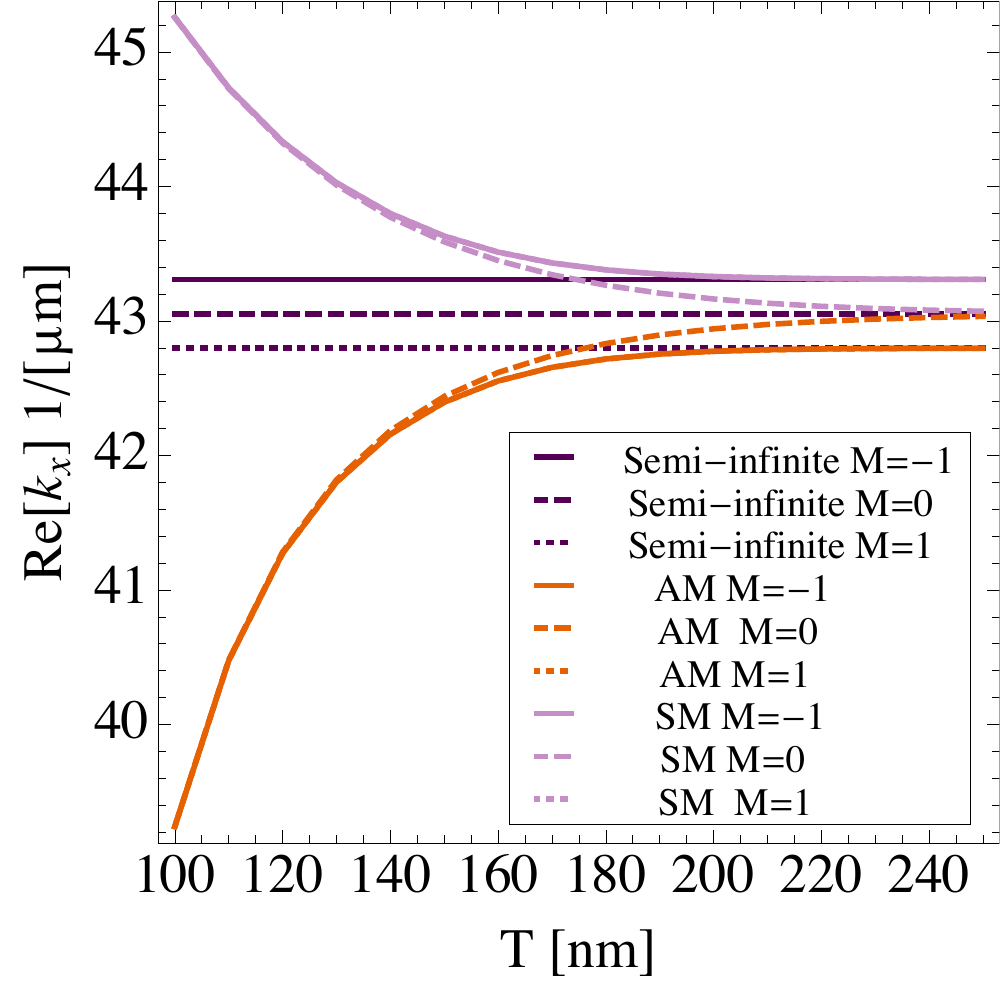} 
\end{subfigure}\\
  \begin{subfigure}[b]{0.47 \textwidth}
(c)  \\
  \centering
 \includegraphics[width= 0.9 \textwidth]{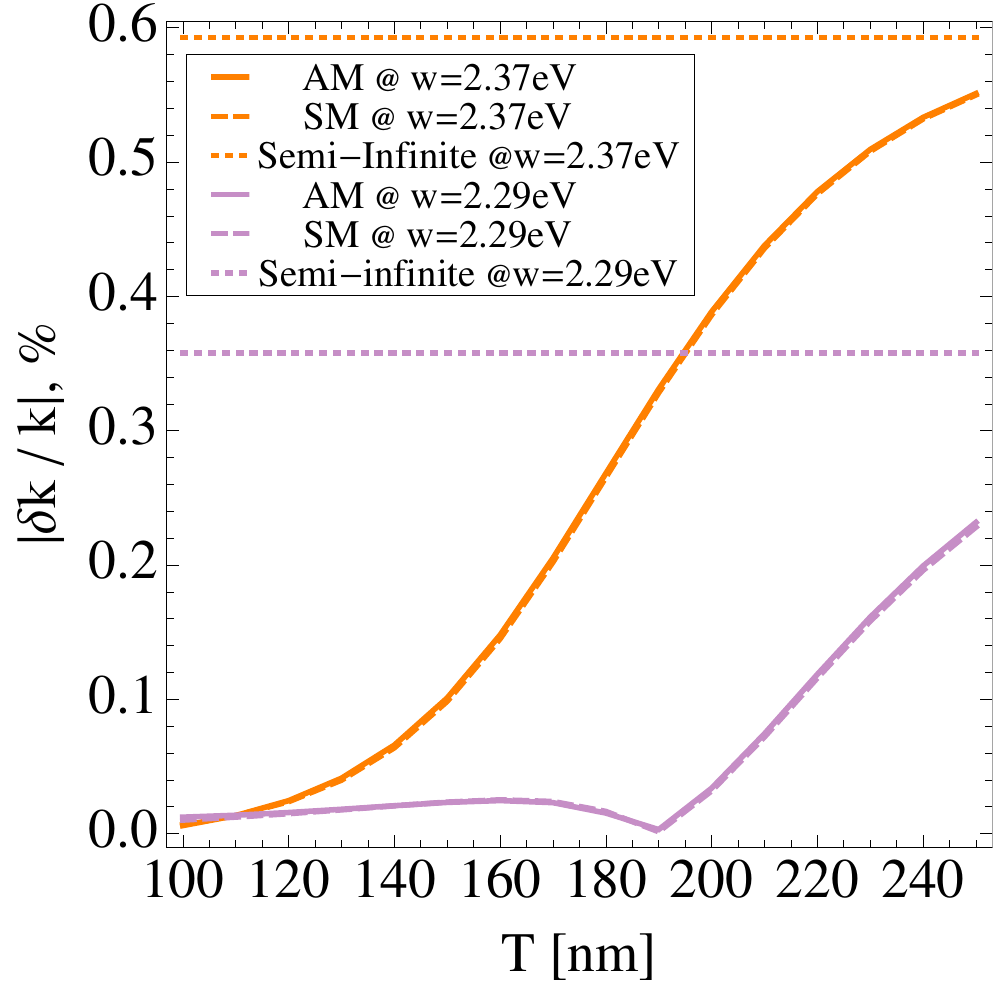} 
\end{subfigure}
\begin{subfigure}[b]{0.47 \textwidth}
(d)   \\
 \centering
 \includegraphics[width=0.95 \textwidth]{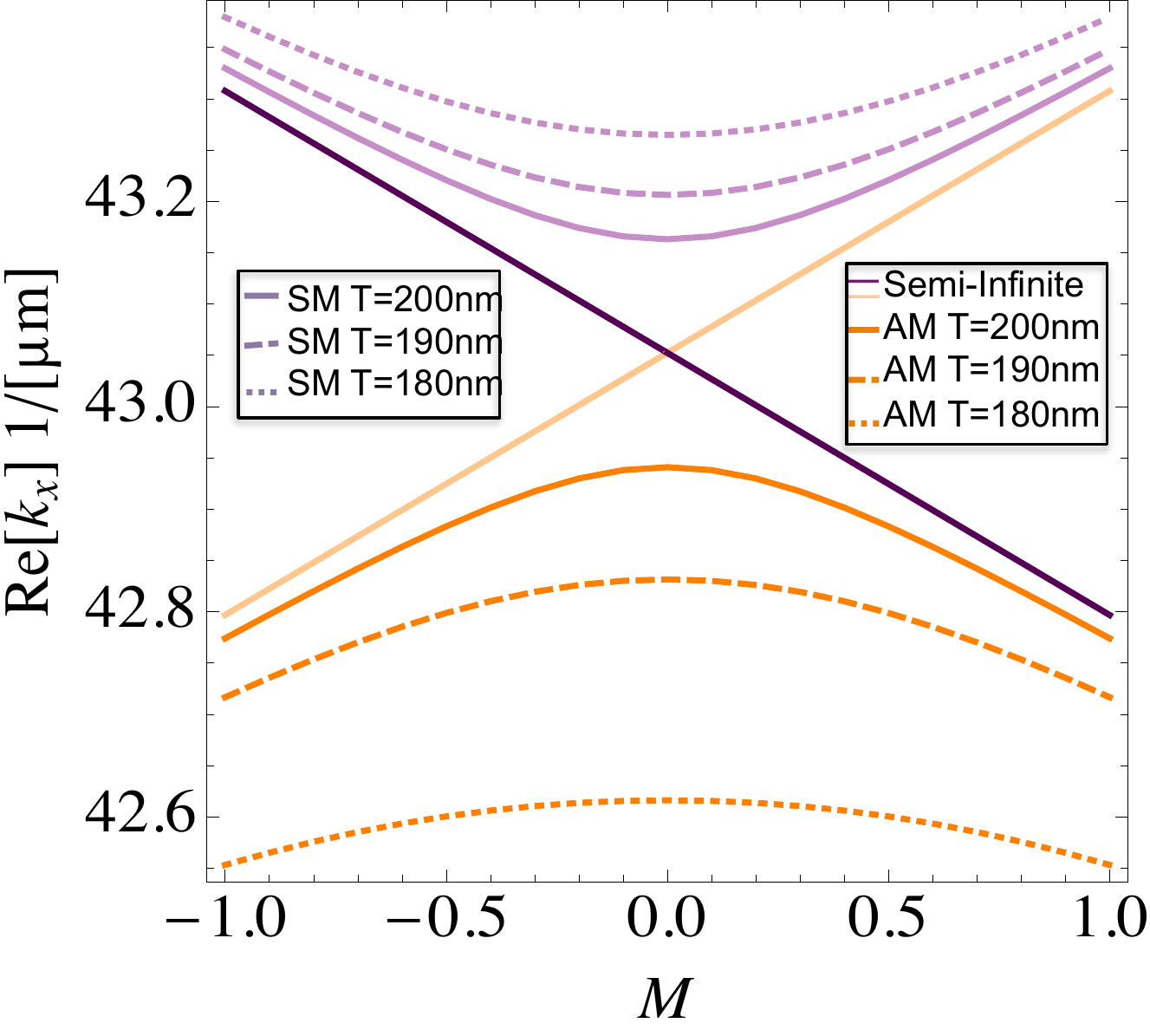} 
\end{subfigure}
\end{tabular}
        \caption{(a) The change in wavevector $\delta k=k_+-k$ resulting from magnetisation of the dielectric as a function of frequency for different cavity thicknesses; (b)  dependence of wavevector on the thickness of the waveguide for $\omega=2.37\,\mathrm{eV}$ and different magnetizations; (c) absolute change in the wavevector vs.  waveguide thickness when the magnetization changes from M=0 to M=-1 for different energies ; (d) dependence of the wavevector on the magnetisation for different cavity thicknesses at $\omega=2.37\,\mathrm{eV}$;  }
   \label{Mdependance}
\end{figure*}

\begin{figure*}[ht]
  \begin{subfigure}[b]{0.47 \textwidth}
 (a)  \\
   \centering
 \includegraphics[width= 0.9\textwidth]{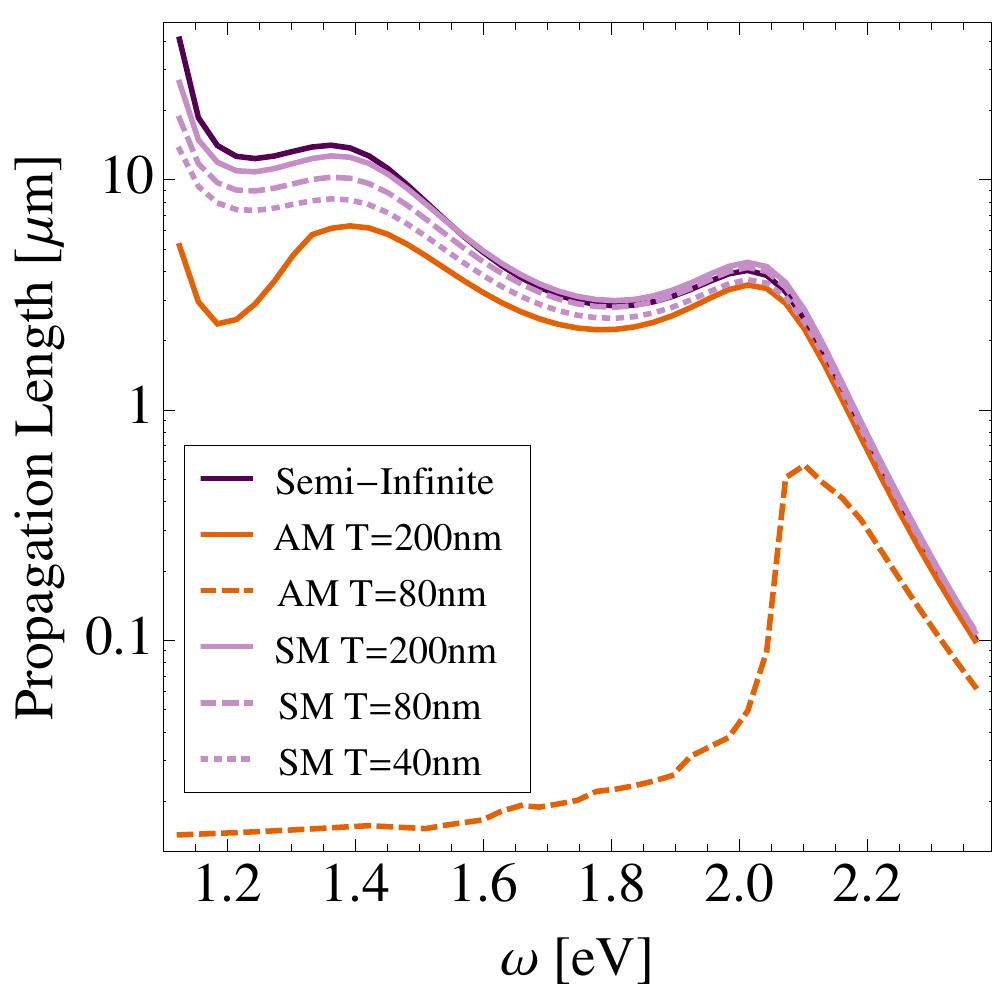} 
\end{subfigure}
   \centering
   \begin{subfigure}[b]{0.47 \textwidth}
   (b)   \\
   \centering
\includegraphics[width= 0.9\textwidth]{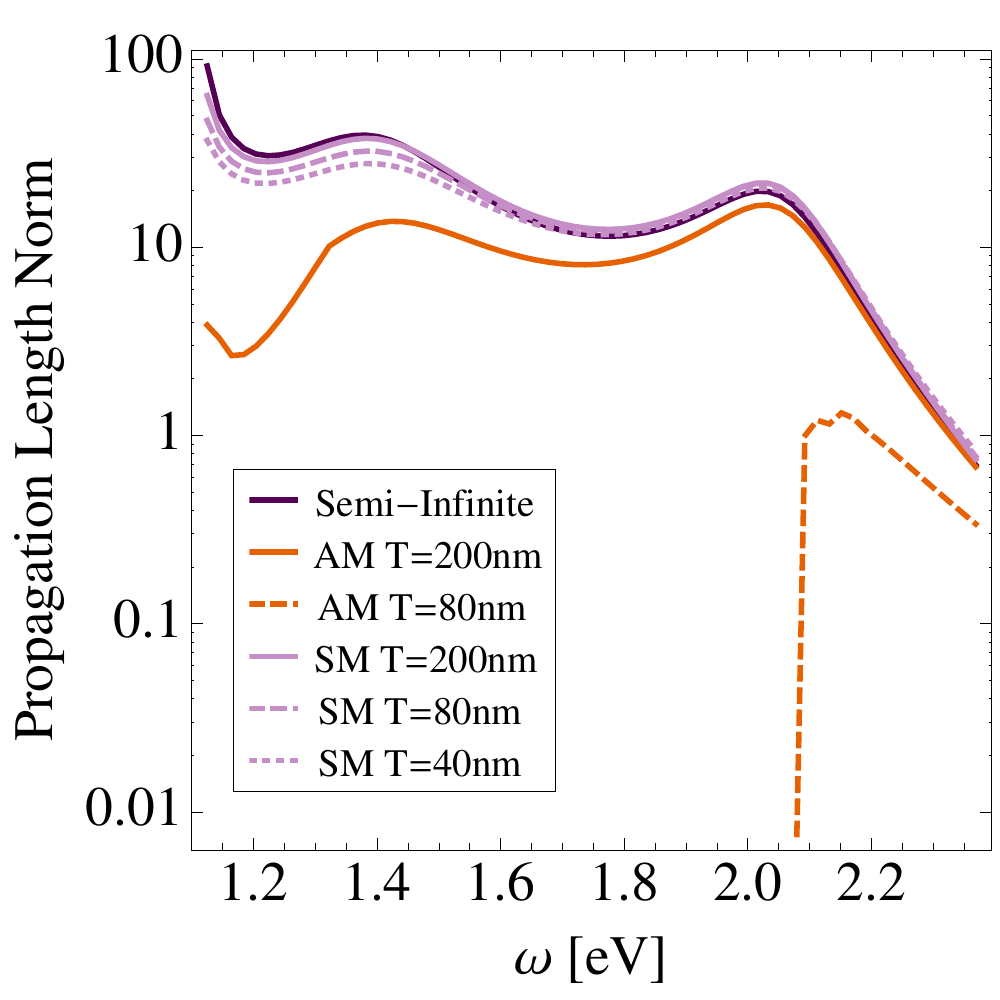} 
\end{subfigure}\\
 \begin{subfigure}[b]{0.47 \textwidth}
 (c)  \\
   \centering
  \includegraphics[width= 0.9\textwidth]{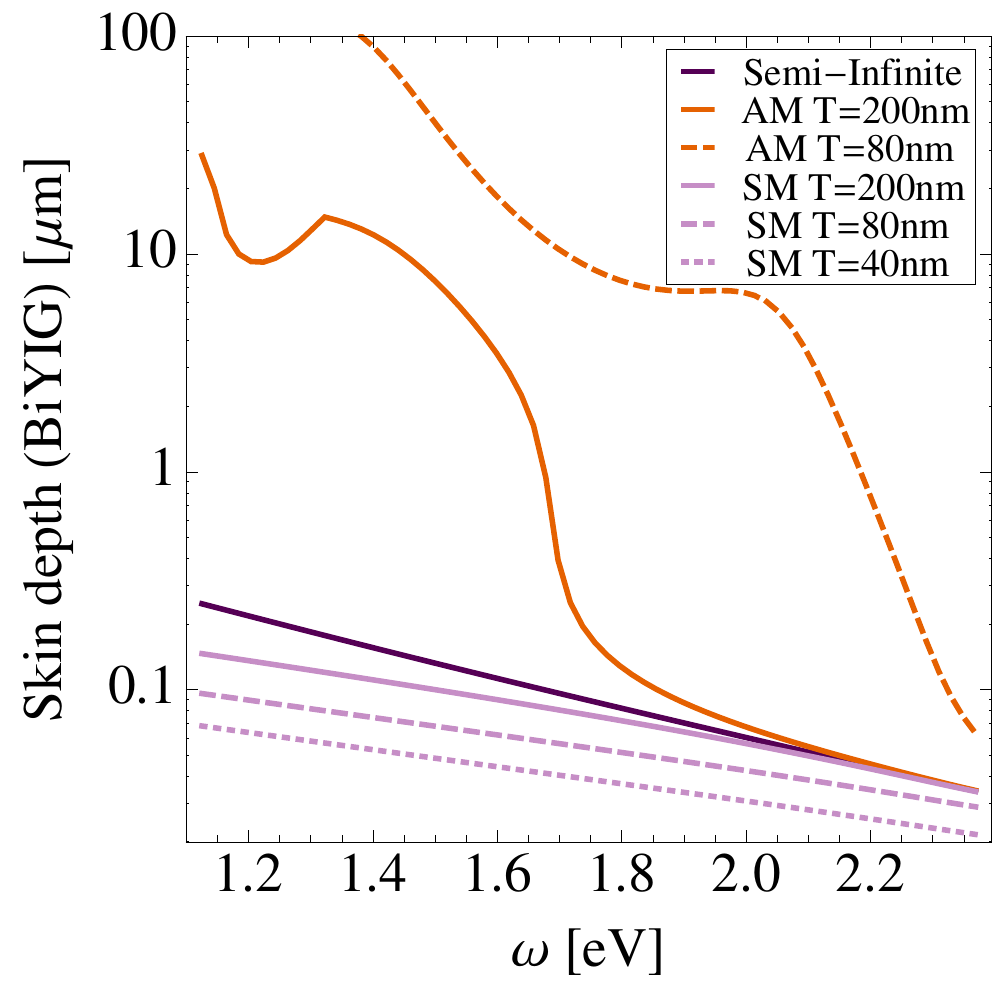} 
\end{subfigure}
 \begin{subfigure}[b]{0.47 \textwidth}
  (d) \\
   \centering
   \includegraphics[width= 0.87\textwidth]{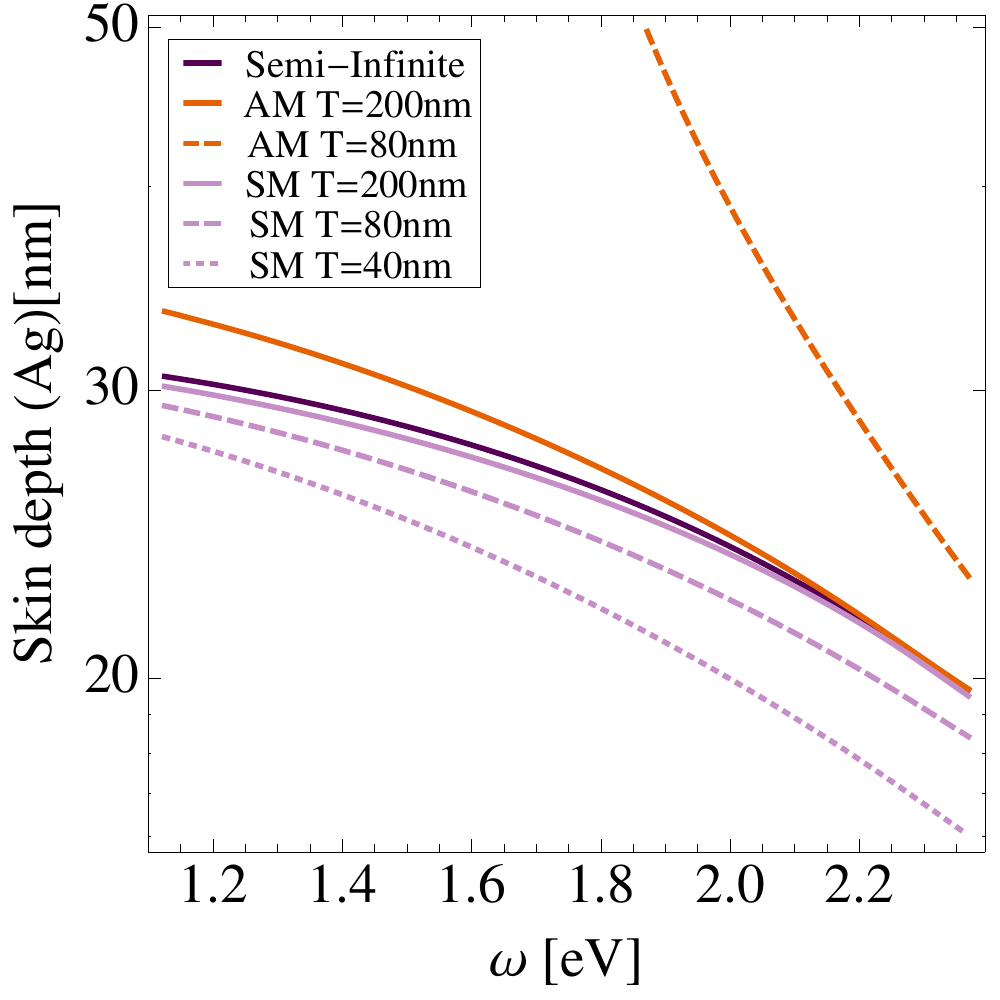} 
\end{subfigure}
   \caption{Decay lengths for a range of cavity sizes (non-magnetic case) as a function of $\omega$:  (a) the propagation length $L_p$; (b) the normalized propagation length $\Re(k_x)L_p/(2\pi)$; (c) the skin depth in the dielectric $\delta_d$; (d) the skin depth in the metal $\delta_m$.}
    \label{Proplength}
\end{figure*}

When a magnetic field is added in the dielectric, the dispersion relation is modified.  We choose a field large enough to saturate the magnetisation, and hence the magneto-optic response; in practice 2\,kOe would be sufficient \cite{BiYIGdata}.  For simplicity we give the magnetisation in units of the saturation value, so magnetisation along $+z$ and $-z$ correspond to $M=\pm1$ respectively.  The changes $\delta k$ in $k_x$ are shown as a function of energy in Figure~\ref{Mdependance}(a). Modest changes (of the order of 1\%) are predicted, the sensitivity to magnetisation being greatest at high energies where the magneto-optical response of the dielectric (Figure \ref{dielectricfns} (b),(d)) is larger. 
For the semi infinite single interface the real part of $\delta k$ depends on the imaginary part of $\epsilon_{xy}$ shown on Figure  \ref{dielectricfns} (d) as can be seen from Eq. (\ref{kxSingle}). (Note that $\epsilon_ m$ is negative hence it is the imaginary part of $\epsilon_{xy}$  which contributes to the real part of $k_x$.) For the cavity $\delta k$ depends on the real part of $\epsilon_{xy}^2$ as seen from Eqs. (\ref{kxSolsSym}) and (\ref{kxSolsAnti}). For large cavities the absolute values of $\delta k$  for the symmetric and antisymmetric modes converge to the same value. 

Figure \ref{Mdependance} (b) shows the wavevector of the propagating modes in the $x$ direction for $\omega =2.37\,\mathrm{eV}$, which is the highest frequency shown on Figure \ref{DispRelation},  for waveguides with different thickness. The wavevectors for the single interface are shown for reference. The wave vector for the cavity modes doesn't depend on the sign of the magnetisation as is clear from the dispersion relation Eq. (\ref{MIMdisp}). For thin waveguides the symmetric and antisymmetric modes are far apart. As the thickness increases the  wavevector approaches the solutions for a single interface. The change in the wavevector versus the waveguide thickness for change in the magnetisation from $M=0$ to $M=-1$ for different energies is shown on figure \ref{Mdependance} (c)  At small thickness the difference in the magnetisation modes is insignificant. The waveguides are less sensitive for the magnetisation at lower energies. This is due to the frequency dependence of $\epsilon_{xy}$ which  increases at higher energies. The change of the wavevector with the magnetisation is shown on figure \ref{Mdependance} (d); note the expected quadratic dependence for small $M$ in the finite waveguides. For single interface the change is linear as is clear from Eq. (\ref{kxSingle}). For cavities the dependence is quadratic. The two modes on the two interfaces couple to produce the anti-crossing. The  magnetic field changes further the levels. For small thickness the change of the wavevector due to the coupling of the two interfaces is predominant. At larger waveguide thickness the coupling between the interfaces weakens which makes the magnetic field effect more pronounced. In the limit of large cavities the solution will follow the one for a single interface.

 \begin{figure}[t] 
   \centering
 \includegraphics[width=0.95\columnwidth]{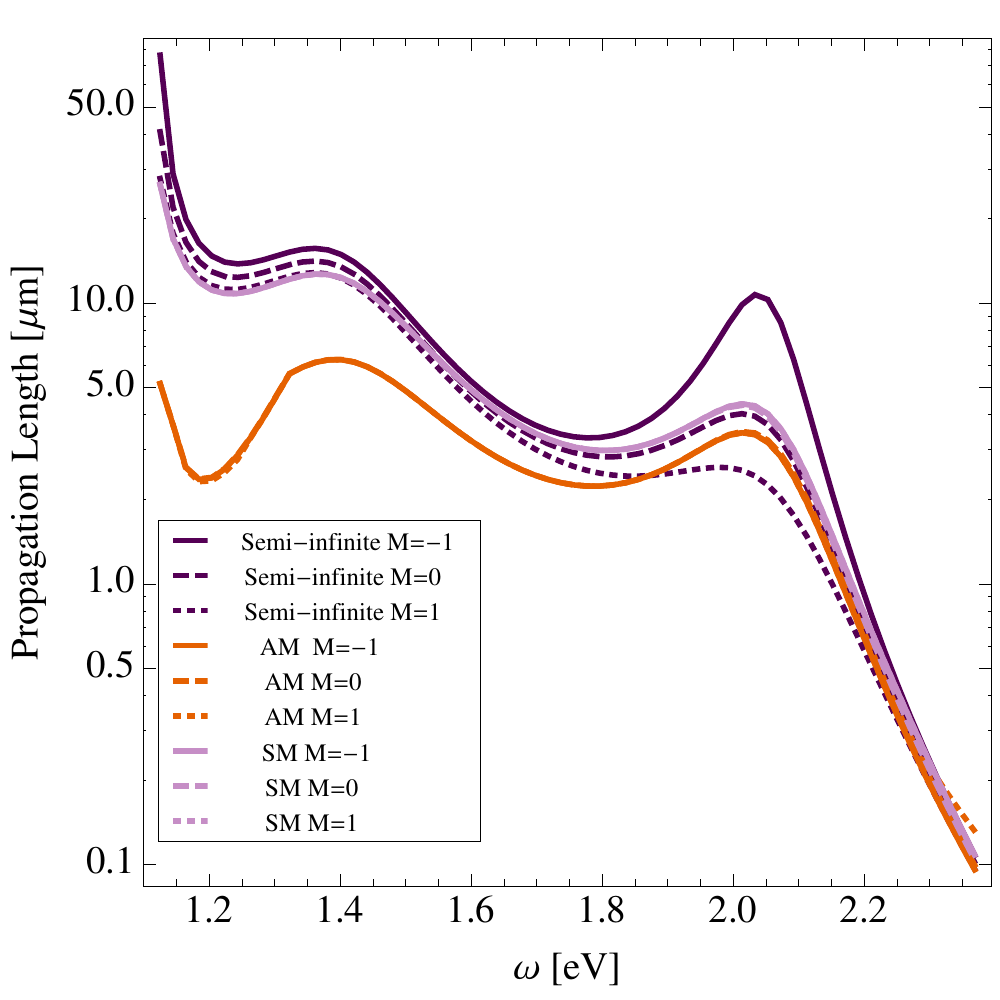} 

   \caption{The  propagation length for cavity of thickness 200\,nm for different magnetizations.}
   \label{PropLengthForM}
\end{figure}

The intensity of a surface plasmon propagating along an interface decays as $\exp(-2 \Im[k_x]x)$, owing to the energy loss from Joule heating; the corresponding propagation length is $L_p=\frac{1}{2\Im[k_x]} $ \cite{Novotny}. Figure \ref{Proplength} (a) shows the propagation length for symmetric and antisymmetric modes for different cavity thickness.   Generally the propagation length rises as the frequency is reduced, especially for modes which approach or cross the dielectric light line so the fields become less strongly bound to the metal-dielectric interfaces;  the plasmons can then propagate long distances of more than 20 $\mu\,\mathrm{m}$.  However, once solutions reach the minimum frequency for propagation along the waveguide without penetrating the electrodes (i.e. approach the frequency axis in Fig.\ref{DispRelation}), the fields are forced to penetrate the metal once again  as the frequency is reduced further and the propagation lengths drop rapidly.  For a given type of mode the propagation length $L_p$ becomes shorter as the cavity thickness $T$ is reduced.

In figure \ref{Proplength}(b) we show the normalised propagation length given by $L_p/\lambda_p$ which is important if we want to apply these waveguides as cavities as it gives the number of oscillation cycles before decay and is therefore a key factor in the determination of the sharpness of the resonance i.e. the quality factor of the cavity. The antisymmetric modes for small cavities have much less than one cycle of oscillation for smaller energies where there are no propagating solutions in the waveguide, in accordance with the dispersion relation. For thicker cavities and symmetric modes there are several cycles of oscillations, but the oscillations decay rapidly for energies above 2 \,eV.   

 Figures \ref{Proplength}  c) and d) show the skin depths in the dielectric core in the metal claddings $\delta_d$ and $\delta_m$. The skin depths are defined as $\delta_d=1/\Re[\kappa_{yd}]$ and  $\delta_m=1/\Re[\kappa_{ym}]$ respectively.  Note that the the smaller skin depth at larger energies of the incident light doesn't correspond to longer propagation length; the losses are instead determined by the proportion of the field energy forced to reside in the metallic regions, which is in turn determined by the geometry of the cavity.

Fig. \ref{PropLengthForM} shows the propagation length for the symmetric and anti-symmetric modes   for a waveguide with a thickness of $200\,\mathrm{nm}$ and for semi-infinite interfaces for different magnetizations. For the symmetric mode the magnetization increases the propagation length while for the anti-symmetric mode it reduces it. This behaviour is implied by Eq. (\ref{kxSols}), which indicates that the magnetisation has different effects on the wavevectors of different modes. 
The difference between the propagation length for different magnetizations for the waveguide modes is insignificant for low energies but becomes larger than the corresponding difference for a single interface at higher energies. 
For a semi-infinite interface the propagation length is different for the positive and negative directions of the magnetisation, as expected from the dispersion relation given by Eq. (\ref{kxSingle}). 

\begin{figure*}[th!] 
     \centering
 \begin{subfigure}[b]{0.3\textwidth}
   \centering
a)  \includegraphics[width=\textwidth]{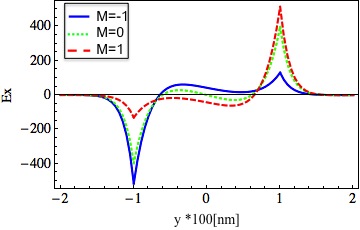} 
\end{subfigure}
  \begin{subfigure}[b]{0.3\textwidth}
   \centering
 b)  \includegraphics[width=\textwidth]{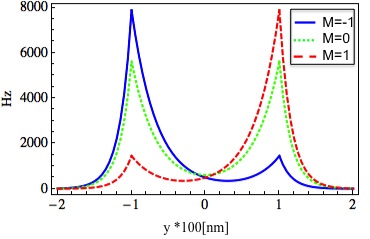} 
\end{subfigure}
   \centering
   \begin{subfigure}[b]{0.3\textwidth}
   \centering
c)   \includegraphics[width=\textwidth]{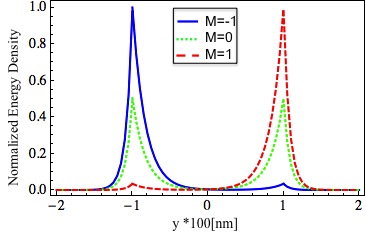} 
\end{subfigure}
   \caption{Field and energy distributions  for symmetric modes as a function of distance from the waveguide median for a Ag/Bi:YIG/Ag structure with core thickness $T=200\,\mathrm{nm}$ at frequency  $\omega=2.37\,\mathrm{eV}$: (a) the electric field $E_x$;   (b) the magnetic field $H_z$; (c) the average total energy density. }
        \label{Fields80nm523}
\end{figure*}

\begin{figure*}[th!] 
     \centering
 \begin{subfigure}[b]{0.3\textwidth}
   \centering
a)  \includegraphics[width=\textwidth]{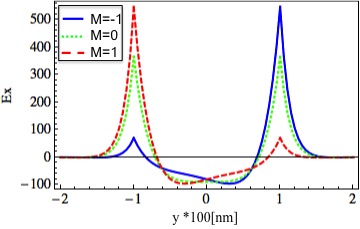} 
\end{subfigure}
  \begin{subfigure}[b]{0.3\textwidth}
   \centering
 b)  \includegraphics[width=\textwidth]{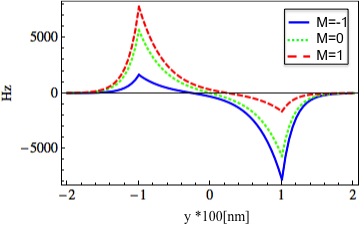} 
\end{subfigure}
   \centering
   \begin{subfigure}[b]{0.3\textwidth}
   \centering
c)   \includegraphics[width=\textwidth]{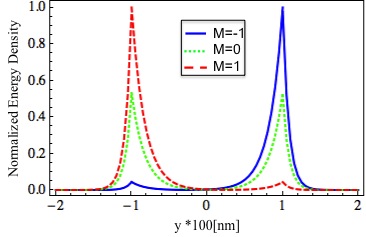} 
\end{subfigure}
   \caption{Field and energy distributions for asymmetric modes as a function of distance from the waveguide median for a Ag/Bi:YIG/Ag structure with core thickness $T=200\,\mathrm{nm}$ at frequency  $\omega=2.37\,\mathrm{eV}$: (a) the electric field $E_x$;   (b) the magnetic field $H_z$; (c) the average total energy density. }
        \label{Fields80nm621}
\end{figure*}

In a non-magnetic plasmonic waveguide the fields have a maximum value at the interfaces and are exponentially decaying away from them. The field has a symmetric or antisymmetric profile inside the waveguide, and the energy density is symmetrically distributed along the two interfaces. In the case of a magnetic waveguide the magnetic field will perturb the symmetry. The field normalizations can be found from the orthogonality condition \cite{ModeOrthogonalityNonreciprocial}
\begin {eqnarray}
\int^{\infty}_{-\infty} dy E_{yn}^\sigma(y) H^\gamma_{m}(y)& -E^\gamma_{ym}(y) H^\sigma_{n}(y) \nonumber \\ 
&=N_{m}^{\gamma,\sigma}\delta_{mn}(1-\delta_{\sigma,\gamma})
\label{Orthogonal}
\end{eqnarray}
where $\sigma,\gamma$ denote the forward and backward propagating waves.  
 The relationship between the fields of the forward and backward propagating modes inside the waveguide are given by
\begin{eqnarray}
E^+_x(M)=E^-_x(-M)\nonumber \\
E^+_y(M)=-E^-_y(-M)
\end{eqnarray}
Hence in Eq. \ref{Orthogonal} the forward and backward propagating fields have to be taken at opposite magnetizations. 
The electric field in the waveguide will be expressed as 
\begin{equation}
\mathcal{ E} = \sum_{n=0}^{M}(a_n\mathbf{E}^+_n(y,z)e^{i k_x^+x}+b_n\mathbf{E}^-_n(y,z)e^{-ik_x^-x})
\end{equation}
where $\mathbf{E}_n^{+,-}$ are the modal profiles in the forward and backward direction satisfying the orthogonality condition Eq. (\ref{Orthogonal}) and $a_n$ and $b_n$ are the mode amplitudes  of the two waves.

The electric and magnetic fields and the energy distribution are shown as a function of $y$ (i.e. distance from the centre line of the dielectric) for the MIM structure in Figure \ref{Fields80nm523} for the symmetric and Figure  \ref{Fields80nm621}   for the anti-symmetric modes.  The mode profiles are obtained from
Eqs.  (\ref{First}) -(\ref{antiE}) and Eq. \ref{Orthogonal}.  The fields shown are calculated at  frequency $\omega= 2.37\,\mathrm{eV}$. At zero magnetisation $M=0$ the fields in Figure~\ref{Fields80nm621} are completely anti-symmetric (for $H_z$) or symmetric  (for $E_x$).  The magnetisation breaks the symmetry and as a result the zero of the $H_z$ field  is shifted from the middle of the waveguide.  The energy density is still  localised along the interfaces but in the presence of magnetisation it is localised preferentially along one of them. The maximum field amplitude is correspondingly different at the two interfaces.  The difference in the field amplitudes depends also on the cavity thickness. The field profiles shown are for a 200 nm thick waveguide. The thinner the waveguide the smaller the difference will be.

\section{Magnetic control of the light field}

 \begin{figure}[hb] 
   \centering
 \includegraphics[width=0.95 \columnwidth]{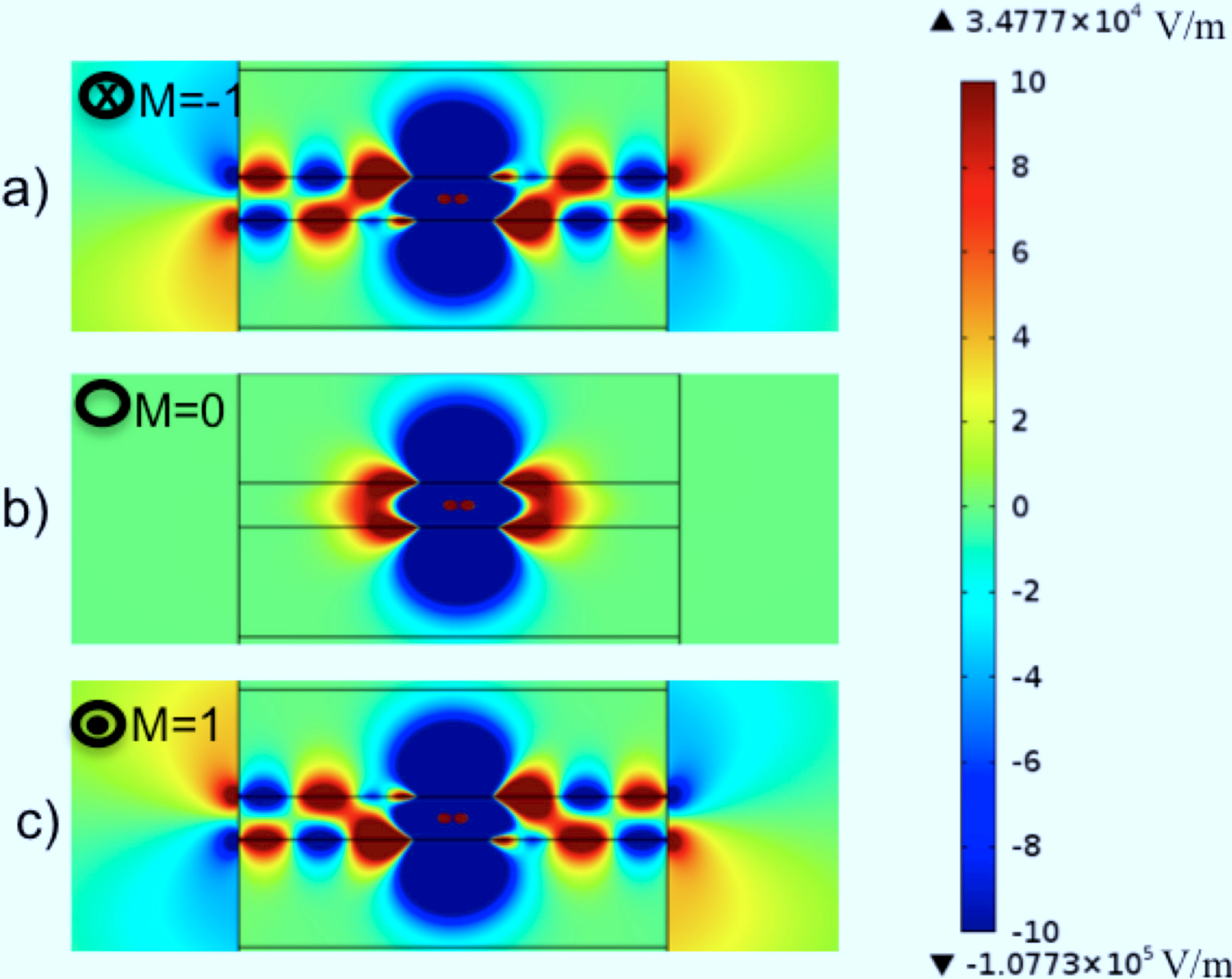} 
   \caption{The $E_x$ field from a radiating electric dipole within a cavity of thickness 40\,nm and length 400\,nm for different magnetisations of the dielectric}
   \label{2DcavityModes}
\end{figure}
 \begin{figure*}[ht] 
   \centering
 \includegraphics[width=0.9 \textwidth]{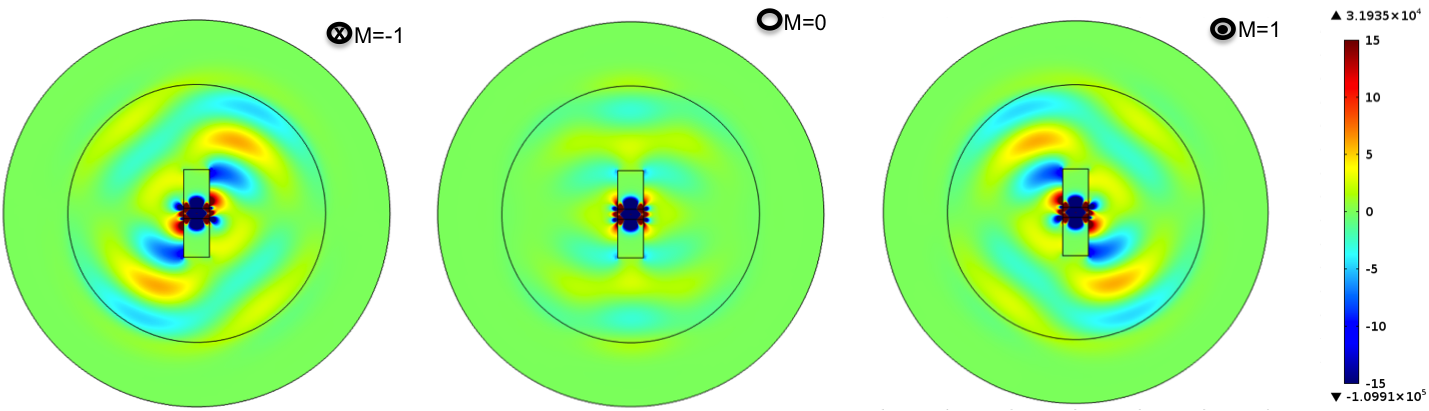} 
   \caption{The near-field distribution of $E_x$ outside a cavity of thickness  80\,nm thick and length 200\,nm containing a radiating dipole at the centre, for different magnetisations of the dielectric}
   \label{DirectionSwitching}
\end{figure*}
To illustrate the effect of these differences, and to show how the relatively small magnetic modifications to the dispersion relation can nevertheless have large effects on the field distributions, we employ finite-element method simulation using commercial software (COMSOL Multiphysics) to find the field distribution in a Bi:YIG magnetic waveguide of 80\,nm thickness and length 400\,nm confined between two silver plates.  We excite the radiation by a point dipole emitter placed in the centre of the structure.  We work at two different excitation frequencies: at the higher frequency  $\omega= 2.37\,\mathrm{eV}$ the waveguide has two propagating  plasmonic modes (one approximately symmetric and one approximately antisymmetric) while at  the lower frequency $\omega=2\,\mathrm{eV}$ there is only one such mode (approximately symmetric).

This shift in the node of the electric field across the cavity with the magnetisation allows us to control the coupling of the dipole to the cavity modes and effectively switch it on and off.   To demonstrate this we simulate  an oscillating electric dipole of magnitude $d$ with a current dipole moment $\omega d=1\,\mathrm{nA}$ and frequency  $\omega= 2.37\,\mathrm{eV} $, oriented along the $x$ axis and placed in the middle  (at coordinates $x=0$ and $y=0$) of a cavity which is  400nm long and 40 nm  thick. Such a narrower cavity supports only one propagating plasmonic mode (symmetric in $H_z$ and anti-symmetric in $E_x$) even at this higher frequency.

Figure \ref{2DcavityModes} shows the $E_x$ field for different magnetizations. At zero magnetisation ($M=0$) the electric field is exactly zero in the middle of the waveguide for the propagating mode, hence this mode is not excited. The electric dipole excites only the antisymmetric (i.e. symmetric in $E_x$) mode which is very rapidly decaying; the resulting field is strongly bound to the central region of the waveguide. When $M\neq 0$, as was shown on Figure \ref{Fields80nm621}, the electric field is also different from zero at the position of the dipole for the symmetric (i.e. antisymmetric in $E_x$) mode; hence both modes are excited. The symmetric mode has a significantly longer propagation length and even though little energy couples into it, it is carried away across the interfaces. The antisymmetric mode decays very fast and at less than half of a plasmon wavelength away from the dipole its electric field becomes equal to that of the symmetric mode.  This leads to a relatively complex interference of the two modes in this region.


These effects also have consequences for the  radiation which couples out of the cavity. We study the radiation emitted into free space for the case where the cavity is 40\,nm thick but only 200\,nm long and where the silver layers are 300\,nm thick. The resulting $E_x$-field is shown on Figure \ref{DirectionSwitching}  for different magnetisations at distances up to $1\,\mathrm{\mu m}$ around the structure.  As can be seen, the magnetic field skews the emitted electric field pattern: the angle of emission is dominated by the geometry of the slot and its effect on the relative phase of the fields around it. 


\begin{figure*}[t] 
     \centering
 \begin{subfigure}[b]{0.32\textwidth}
   \centering
a)  \includegraphics[width=\textwidth]{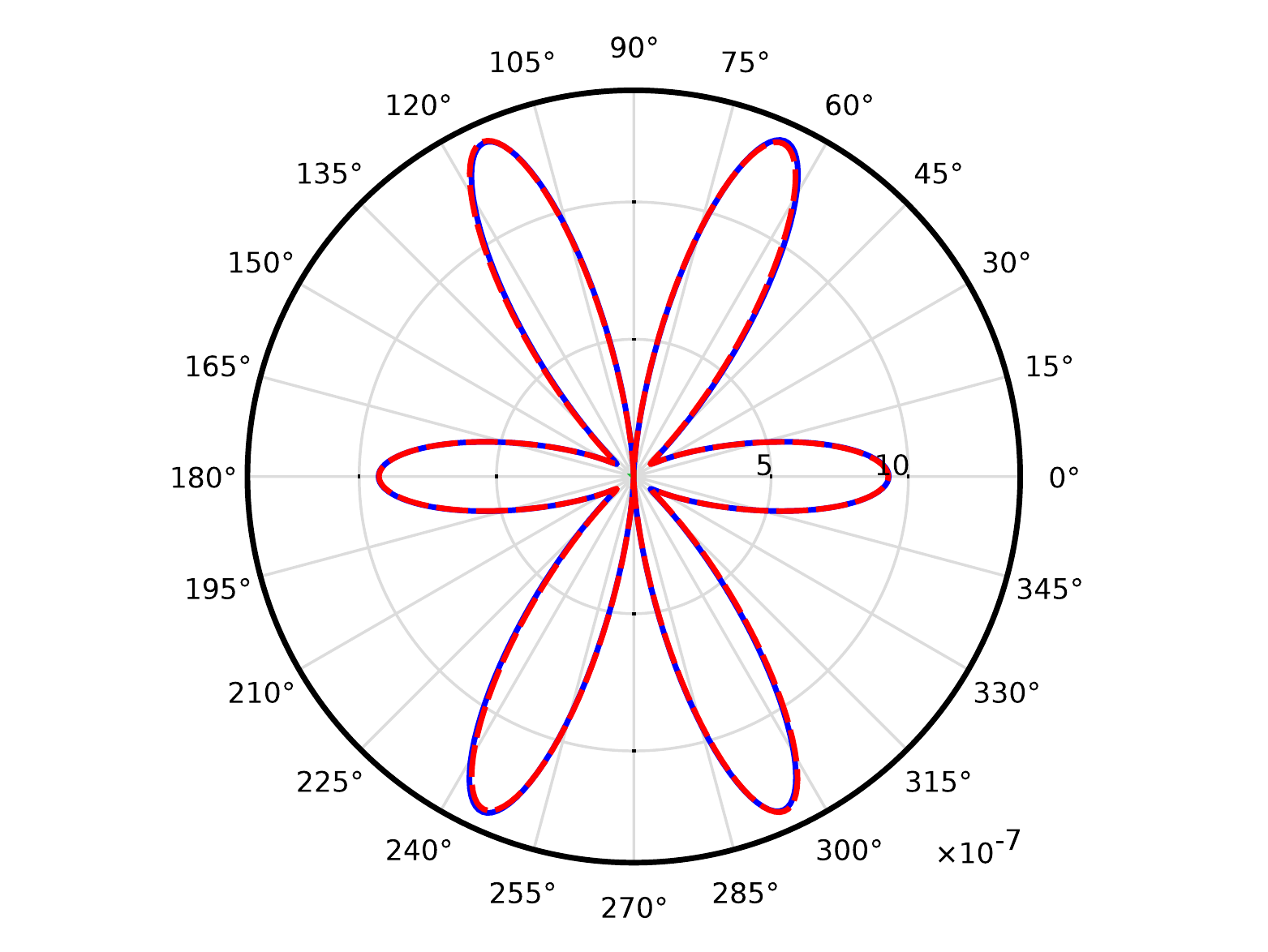} 
\end{subfigure}
  \begin{subfigure}[b]{0.32\textwidth}
   \centering
 b)  \includegraphics[width=\textwidth]{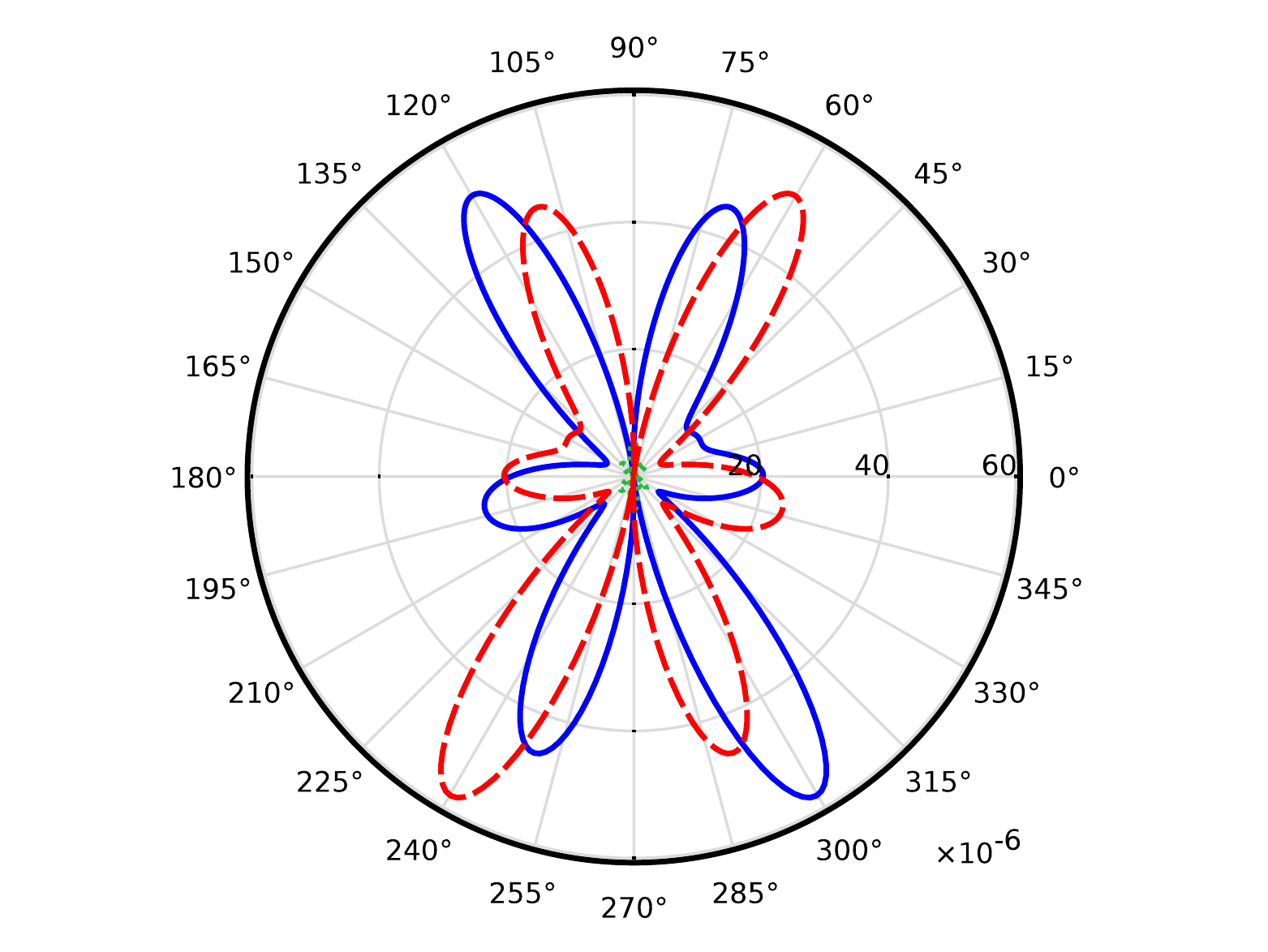} 
\end{subfigure}
  \begin{subfigure}[b]{0.32\textwidth}
   \centering
 c)  \includegraphics[width=\textwidth]{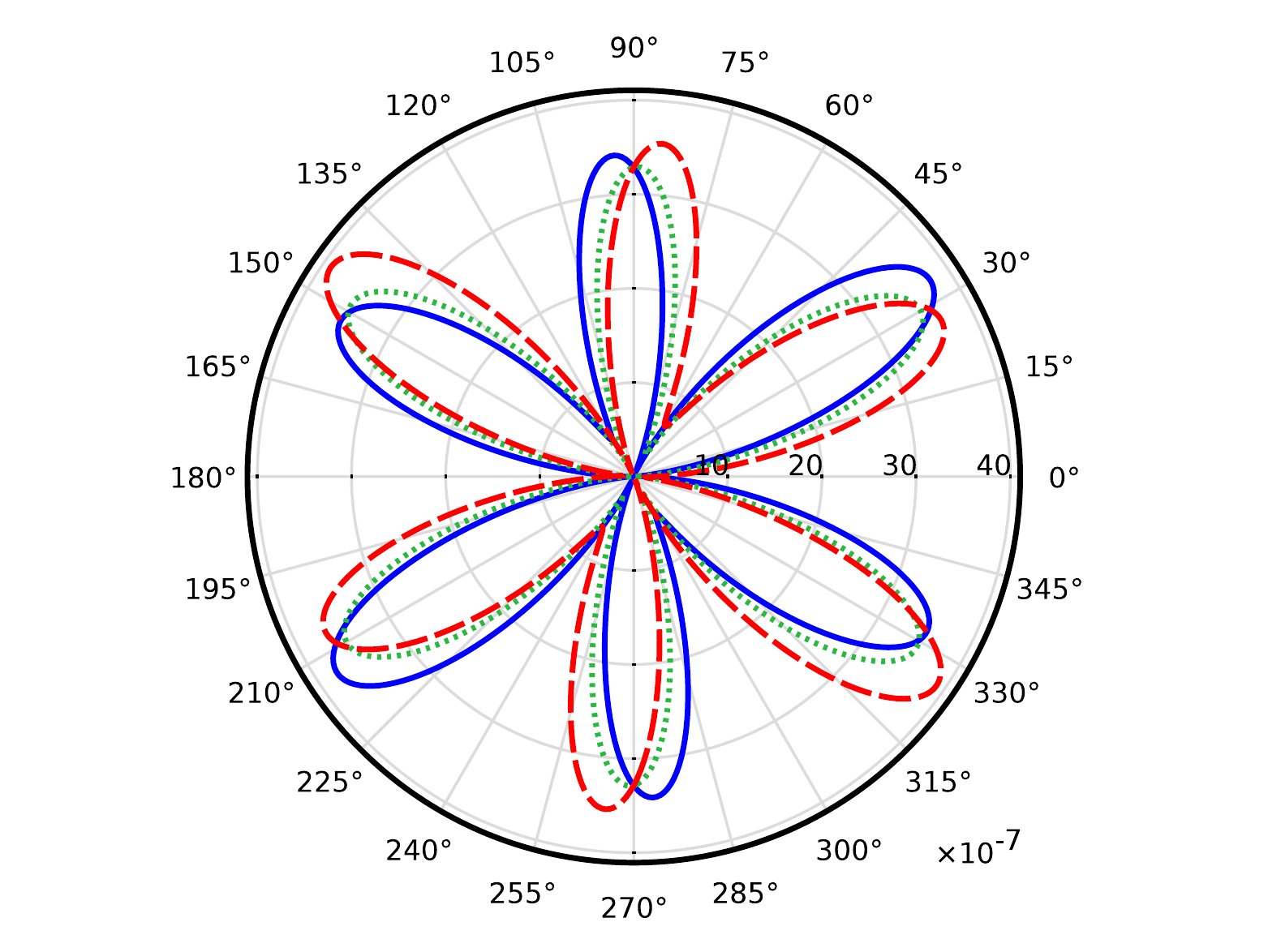} 
\end{subfigure}
\\
   \centering
   \begin{subfigure}[b]{0.32\textwidth}
   \centering
d)   \includegraphics[width=\textwidth]{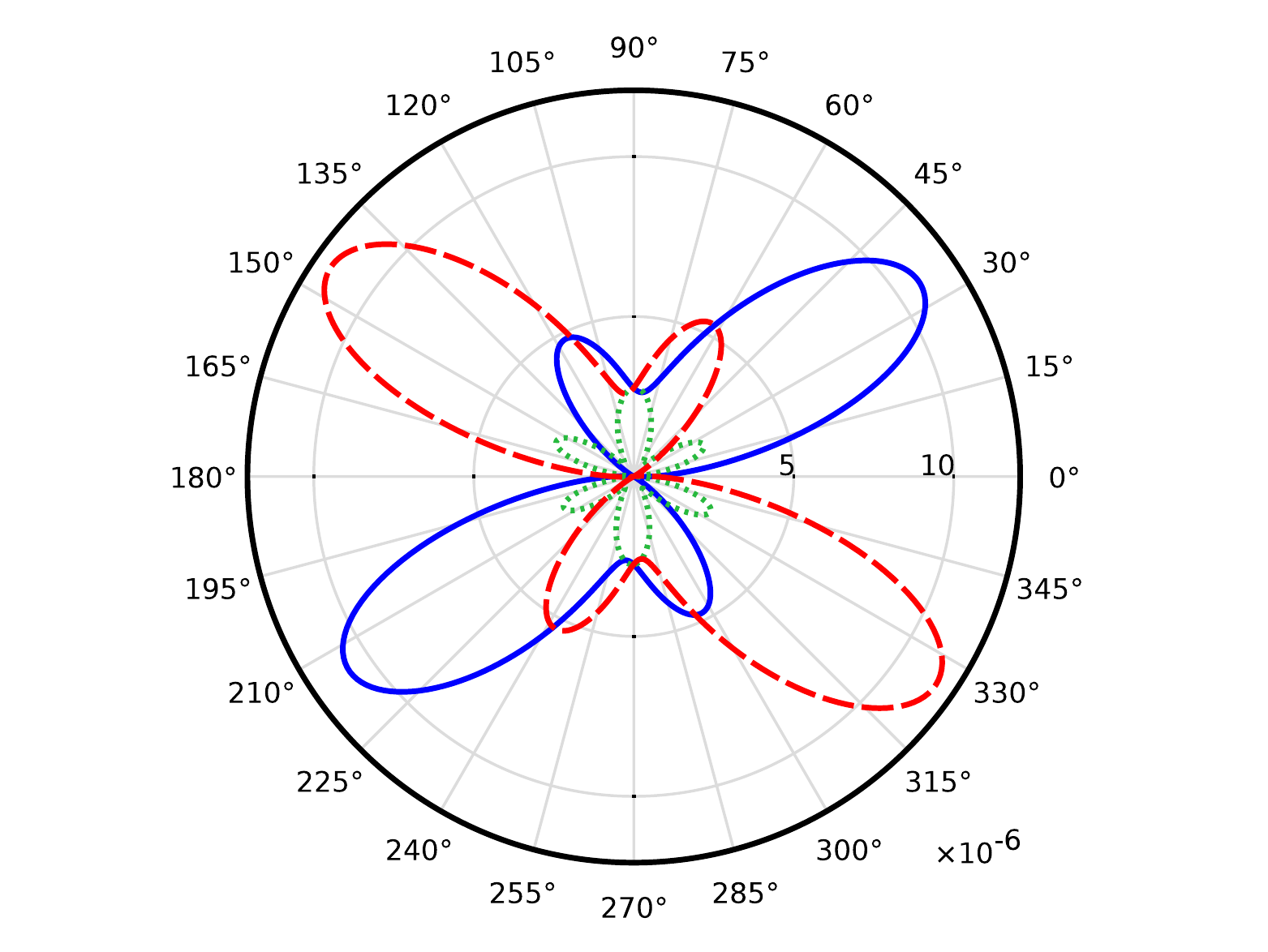} 
\end{subfigure}
 \centering
   \begin{subfigure}[b]{0.32\textwidth}
   \centering
e)   \includegraphics[width=\textwidth]{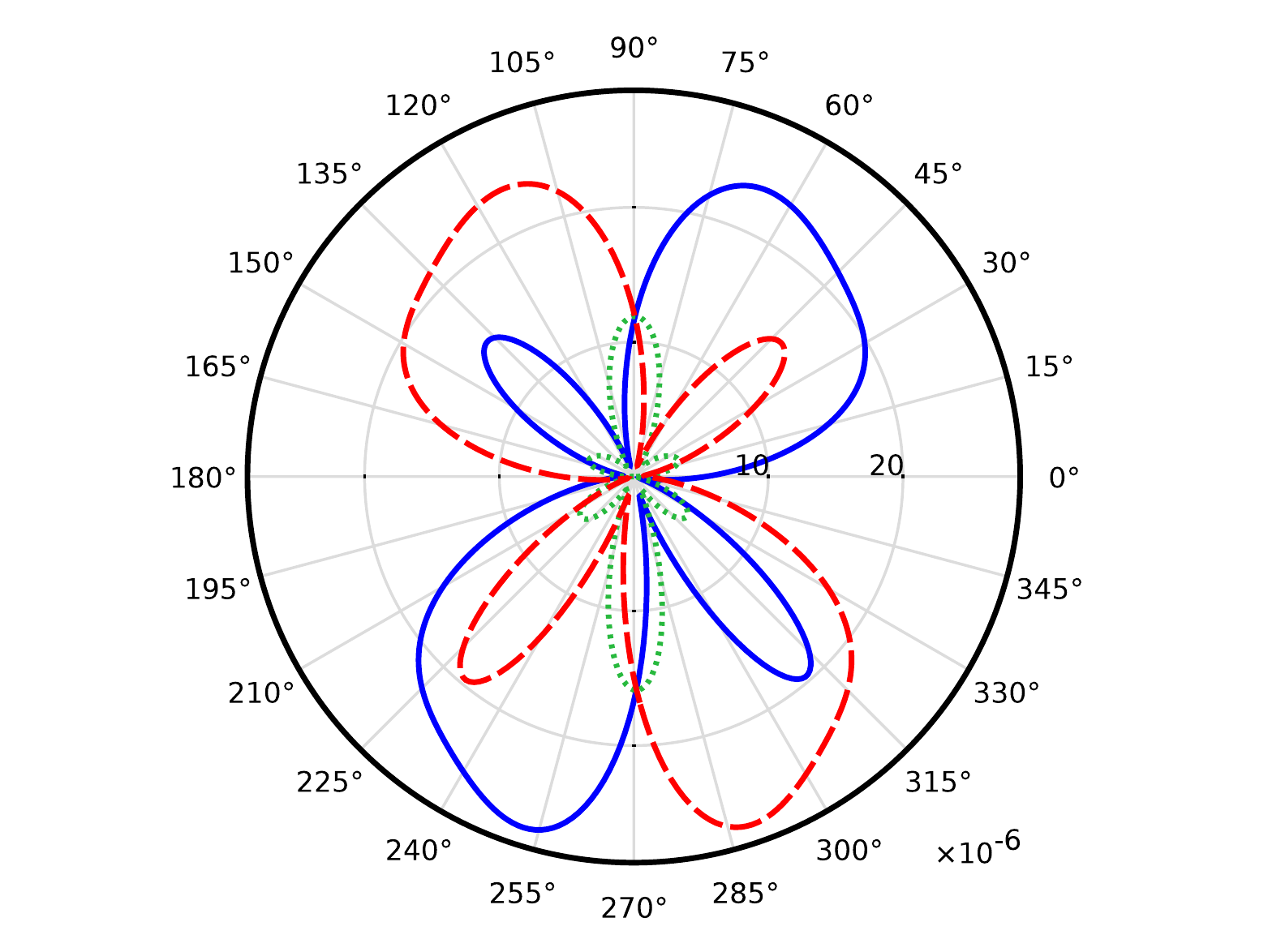} 
\end{subfigure}
  \begin{subfigure}[b]{0.32\textwidth}
   \centering
 f)  \includegraphics[width=\textwidth]{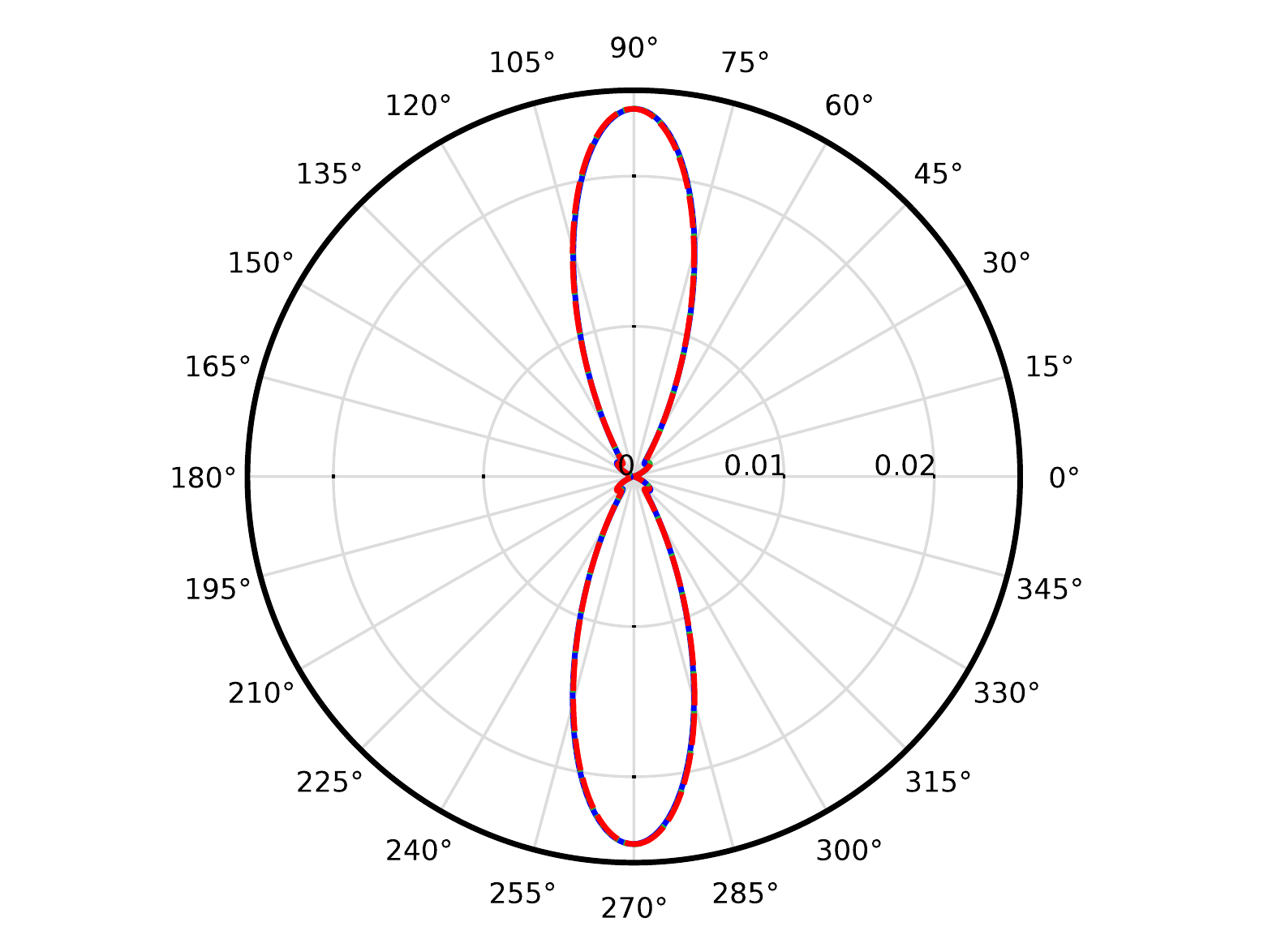} 
\end{subfigure}
   \caption{Polar plots of the far-field radiation intensity $|E|^2=E_x^2+E_y^2+E_z^2$ from a dipole placed in the centre of a cavity filled with magnetic or non-magnetic Bi:YIG. The magnetic state of the dielectric is given by the colour or the curves(blue for M=-1, dotted green for M=0 and dashed red for M=1): (a) for a cavity of thickness 40\,nm and width 400\,nm at $\omega= 2.37\,\mathrm{eV} $ (b)  for a cavity of thickness 80\,nm and width 400\,nm at $\omega= 2.37\,\mathrm{eV} $ (c)  for a cavity of thickness 80\,nm and width 400\,nm $\omega=2\,\mathrm{eV}$.   (d) for a cavity of thickness 40\,nm and width 200\,nm  at $\omega= 2.37\,\mathrm{eV} $ (e)  for a cavity of thickness 80\,nm and width 200\,nm at $\omega= 2.37\,\mathrm{eV} $ (f)  for a Bi:YIG slab with thickness 680\,nm and width 400\,nm (no cavity) at $\omega= 2.37\,\mathrm{eV} $}
    \label{RadiationDipole}
\end{figure*}

In Figure \ref{RadiationDipole} we show the emitted radiation in the far field, showing several different geometries in order to illustrate the different regimes.  The first three subfigures are for a 400nm long cavity. At $\omega=2.37\,\mathrm{eV}$ (figures  \ref{RadiationDipole} a) and b) the radiation at zero magnetisation is largely suppressed, since (as shown previously) no energy couples to the symmetric propagating mode (anti-symmetric in $E_x$),  while the antisymmetric mode has very short propagation length. In the presence of magnetisation the electric field distribution is only slightly different for different magnetisation; the main role of $M$ is to break the symmetry and turn on the radiation. For shorter cavities, shown in figures  \ref{RadiationDipole} d) and e), the radiation emitted into free space at zero magnetisation is due to the energy coupled to the antisymmetric mode. When the magnetisation is different from zero the two modes are both excited; although they have different degrees of excitation in the centre of the structure, they decay so that the field amplitudes at the edge are comparable.  The magnetic field then tunes the relative phases of these contributions and alters the interference from constructive to destructive, depending on the direction.  The resulting interference pattern of the far- field radiation is therefore skewed to different sides depending on the magnetisation. 
 For 80nm cavity with a dipole oscillating at $\omega=2\,\mathrm{eV}$ in short cavities such as the 400\,nm (figure  \ref{RadiationDipole} c) )  the antisymmetric mode has not decayed by the end of the cavity; hence the radiation patterns at $M=0$ and for $M\ne0$ are similar. Finally we show in Figure \ref{RadiationDipole}(f) the radiation from a dipole placed in the middle of a nanostructure consisting only of a slab of magnetic insulator, with no metal cavity.  Here a change in the magnetic field does not result in any noticeable change in the far-field radiation; this shows that the observed effects are indeed due to the surface plasmons.

\section{Conclusion}

The modes of a magnetised MIM waveguide form an interesting contrast to the modes of a single magnetic interface.  Because the system as a whole has a plane of symmetry at $y=0$,  the allowed values of the wavenumber are not direction dependent for a given frequency.  Instead, the modes come in pairs with very different weights of the electric field on the upper and lower waveguide surfaces.  


Even though the changes in the dielectric function introduced by the presence of the magnetisation may be small in absolute terms, these qualitative changes in the nature of the optical modes can lead to significant changes in the response of the structure to excitation.  We have illustrated this by computing the field distributions in a finite-length slot waveguide in response to excitation by a point dipole placed at its centre.  Both the total emission from the cavity and the pattern of far-field radiation can be strongly modified by switching the magnetic field; the total emission is predominantly controlled by the presence or absence of propagating modes with strong coupling to the radiating dipole, while the far-field radiation pattern is controlled by the phase differences at the opposite ends of the guide.  These examples point to the possibility of using magnetic control to switch the propagation of fields in more complex photonics structures.

One possible limitation in applications comes from the match between the properties of the magnetic dielectric and of the metal used in the waveguide.  For the materials used here, the magnetic effects become largest above $\omega\approx2\,\mathrm{eV}$, where the silver layers remain weakly dissipative.  If another metal, for example gold, were used that had a strongly dissipative response above 2\,eV, shorter structures would be needed for the effects we describe to survive.

\begin{acknowledgments}
We acknowledge the support from the European Commission Marie-Curie Fellowship for the TASMANIA project.  We thank Markus Schmidt for useful discussions and for communicating the numerical tables for the dielectric tensor of Bi:YIG.
\end{acknowledgments}

\end{document}